\documentclass[a4paper,12pt]{article}

\usepackage{bbm}
\usepackage[utf8]{inputenc}
\usepackage{placeins}
\usepackage{amssymb,amsmath,amsfonts}
\usepackage[colorlinks,linkcolor=purple,citecolor=teal]{hyperref}
\usepackage{dsfont}
\usepackage{color}
\usepackage{graphicx}
\usepackage{hyphenat}
\usepackage{wrapfig}
\usepackage{empheq}
\usepackage{textcomp}
\usepackage[caption=false]{subfig}
\usepackage{rotating}
\usepackage{wrapfig}
\usepackage{pdfpages}
\usepackage{verbatim}
\usepackage{setspace}
\usepackage{array}
\usepackage{caption}
\usepackage[pdf]{pstricks}
\usepackage{upgreek}
\usepackage{cmll}
\usepackage{latexsym}
\usepackage{braket}
\usepackage{epsfig}
\usepackage[left=2.3cm,top=3.2cm,right=2.3cm,bottom=3.2cm,bindingoffset=0cm]{geometry}
\usepackage[titletoc,page]{appendix}
\usepackage{multicol}
\usepackage{youngtab}
\usepackage{enumerate}

\usepackage[backend=bibtex,style=numeric-comp,sorting=none,natbib=true, url=false, doi=false, isbn=false]{biblatex} %

\usepackage{tikz-cd}

\usepackage{tcolorbox}
\tcbuselibrary{breakable}

\newcommand{\beq}{\begin{eqnarray}}
\newcommand{\eeq}{\end{eqnarray}}
\newcommand{\bea}{\begin{eqnarray}}
\newcommand{\eea}{\end{eqnarray}}
\newcommand{\be}{\begin{equation}}
\newcommand{\ee}{\end{equation}}

\newcommand{\im}{\mathrm{i}}

\newcommand{\dd}{\mathop{\mathrm{d}\!}{}}
\newcommand{\deriv}[2]{\dfrac{\dd #1}{\dd #2}}

\def\brc{\langle }
\def\ckt{\rangle}

\def\nn{\nonumber}

\numberwithin{equation}{section}

\def\l{{\tilde{\ell}}}

\numberwithin{equation}{section}

\begin{document}

\title{
\vskip 20pt
\bf{   The Monopole-Fermion Problem \\  in a %
Chiral Gauge Theory }  
}
\vskip 40pt 
\vskip 40pt   
\author{
	Stefano Bolognesi$^{1,2}$, Bruno Bucciotti$^{3,2}$ and Andrea Luzio$^{3,2}$
\\[13pt]
{\em \footnotesize
$^{(1)}$Department of Physics ``E. Fermi", University of Pisa}\\[-5pt]
{\em \footnotesize
Largo Pontecorvo, 3, Ed. C, 56127 Pisa, Italy}\\[2pt]
{\em \footnotesize
$^{(2)}$INFN, Sezione di Pisa,    
Largo Pontecorvo, 3, Ed. C, 56127 Pisa, Italy}\\[2pt] 
{\em \footnotesize
$^{(3)}$Scuola Normale Superiore,   
Piazza dei Cavalieri, 7,  56127  Pisa, Italy}\\[2pt]
{ \footnotesize  stefano.bolognesi@unipi.it, \ \  bruno.bucciotti@sns.it,  \ \  andrea.luzio@sns.it}
}

\date{}

\vskip 8pt
\maketitle

\begin{abstract}

The scattering of electrically charged fermions on magnetic monopole leads to the Callan-Rubakov problem. We discuss some aspects of this problem for Abelian gauge theories with chiral fermions in a Dirac monopole background. In some cases, it is possible to embed the theory in a non-Abelian gauge theory where the monopole is regularized as a 't Hooft-Polyakov monopole. One theory of this kind is the  $SU(N)$ chiral gauge theory with fermions in the representation $\tiny{ \yng(2) \oplus  {\bar  {\yng(1,1)}}  \oplus   8  \times  {\bar  {\yng(1)}}}$, also called ``$\psi \chi \eta$'' model,  with an extra adjoint scalar that induces the Abelianization of the gauge group. We examine this model in detail and provide a possible solution for the condensates around the monopole,  the symmetry preserving boundary conditions, and discuss the particle scattering problem.

\end{abstract}

\newpage
\tableofcontents

\section{Introduction}

Magnetic monopoles have been rigorously studied for almost 100 years. Therefore it is somehow surprising that the problem of scattering massless charged fermions on magnetic monopoles still leads to an unsolved puzzle. For certain theories one cannot find, among the low-energy outgoing states, one that reproduces all the conserved quantum numbers of the incoming particle. This problem is sometimes called the unitarity (or missing final state) puzzle. The class of theories that have this problem includes vectorial models with multiple flavors of massless Dirac fermions, as well as the even less explored chiral theories.

The original problem is already present if one considers the free propagation of a charged Weyl fermion on the static Dirac monopole background \cite{Kazama:1976fm}. While an incoming fermion with non-minimal angular momentum feels a centrifugal barrier and bounces on the monopole, never reaching its core, the solution of the Weyl equation with minimal angular momentum (depending on the fermion charge) is a purely in-going or out-going wave that arrives straight at the monopole core.
This renders the outcome of the scattering not determined by the low-energy effective field theory (EFT) alone, as aspects of the UV completion are relevant. A vast literature focuses on the case in which the UV completion is a vectorial theory\cite{Callan:1982ah,Callan:1982au,Callan:1982ac,Callan:1983tm,Rubakov:1982fp,Rubakov:1988aq,Kazama:1976fm,Polchinski:1984uw,Affleck:1993np,Sen:1984qe,Shnir:2005vvi}, and for the simplest models concrete computations lead to surprising predictions, such as the presence of a chiral symmetry breaking condensate around the monopole \cite{Rubakov:1982fp,Rubakov:1988aq,Kazama:1976fm} (which we dub as Callan-Rubakov condensate from here on), that leads to the baryon decay catalysis in some GUT models.

One might expect that considering the UV completion should also solve the scattering problem. On the contrary, the concrete computations confirm that in many theories (e.g. QED with $N_f\ge 4$ (Dirac) flavors\footnote{Once regularized as a $SU(2)$ Yang-Mills theory with $N_f$ (Weyl) flavor of fundamental fermions.}) it is impossible to reproduce the quantum number of the in-going state with a state built of out-going fermionic modes, with the condensate failing to break the symmetry group enough to allow for consistent final scattering states. The S-matrix applied to an IN state gives an OUT state with fractional numbers of particles whose physical interpretation is still not clear. The fractional number in the $N_f =4$ case is $1/2$  and  these states also sometimes called ``semitons".  Notably some of the approaches, starting from Callan's seminal works \cite{Callan:1982ah,Callan:1982au,Callan:1982ac,Callan:1983tm}, provide a putative solution which however suffers from conceptual problems, so it cannot be fully accepted.

Despite many recent efforts, a precise and satisfying description of the outcome of scattering is still lacking, with many discordant proposals competing to give the correct solution  \cite{Hamada:2022eiv,vanBeest:2023mbs,
	vanBeest:2023dbu,Csaki:2020yei,
	Csaki:2020inw,
	Csaki:2021ozp,Brennan:2021ewu,	Csaki:2022qtz,Khoze:2023kiu,Kitano:2021pwt,Brennan:2023tae,Csaki:2024ajo}.

Many of these analyses do not start from a UV complete model, but directly from the low-energy effective description. This approach has the advantage that the details of the models are abstracted away. However, there is also the risk of trying to solve an impossible problem, where the effective description does not admit a UV completion thus it is not guaranteed that a consistent solution to scattering even exists.

The problem with this approach is possibly more serious when applied to chiral gauge theories as the solution of the puzzle might be different, e.g. if the Callan-Rubakov condensate breaks a bigger part of the global symmetry group, or higgses part of the Abelian symmetries. Despite these logical possibilities, a complete study of the Callan-Rubakov effect in chiral gauge theories is lacking in the literature.

Within this paper, we will not try to address directly the unitarity puzzle in a chiral gauge theory. Instead, we aim to take the preliminary step where, from the UV completion, one determines the symmetries preserved by the system and the correct effective boundary condition for the effective degrees of freedom.

In particular, in the vector-like examples, the  works of Callan and Rubakov can be summarized and subdivided into several steps:
\begin{enumerate}[i]
	\item First, one starts from a consistent UV complete model, where the Dirac monopole is realized as a smooth 't Hooft-Polyakov monopole. 
	\item Then one works out the IR $4$D effective theory, that describes well the low energy theory far away from the monopole core.
	\item Exploiting the rotational symmetry of the Dirac monopole, one decomposes the $4$D degrees of freedom into their angular momentum towers of modes. 
	\item One realizes that the higher angular momentum modes are unimportant in describing the physics of the lowest angular momentum modes at low energy and weak electric coupling, as they scatter far away from the monopole and decouple form the interesting physics.\footnote{Higher angular momentum modes are coupled to the lowest angular momentum modes by exchange of higher angular momentum modes of the gauge field fluctuation and finite size effect of the monopole core. Nevertheless, these effects should be small corrections on top of the zero-order solution. For an analysis of the first effect see \cite{Rubakov:1982fp,Rubakov:1988aq}, while we estimate the latter in appendix A} This leads to an effective $1+1$ dimensional description where the important degrees of freedom are the lowest angular momentum modes (``s-wave" modes) of fermions and the abelian gauge field. Importantly the UV completion of the Dirac monopole as a 't Hooft-Polyakov monopole fixes the boundary condition of the $r=0$ worldline.
	\item Then one can compute the Callan-Rubakov condensate that breaks the anomalous symmetries around the monopole core, allowing for some processes that are impossible in the bulk, far away from the monopole core.
	\item Equivalently, in the 2D EFT that describes the ``s-wave" modes, one can bosonize the fermion fields and integrate away the gauge field, obtaining a new effective boundary condition that captures all the dynamics due to the monopole. One is left with a free fermion theory on $\mathbbm R \times \mathbbm R^+$ ($r>0$), with a fixed boundary condition on the monopole location $r= 0$.
\end{enumerate}
We aim to perform a similar analysis in a chiral gauge theory, giving a more solid base where to start a discussion on the scattering problem.

In doing so, we notice that by starting from a (possibly chiral) gauge theory based on a semi-simple gauge group, and breaking it to its Cartan, the low energy theory is necessarily \textit{locally vector-like}, meaning that there exists a suitable basis of the Cartan such that the spectrum is vector-like under each element of the basis individually. This poses a serious constraint on the space of the chiral IR model we can consider if we wish to maintain the entire Cartan intact. Among the models we can study, one of the simplest is an Abelian quiver theory which can be UV-completed to the $\psi\chi\eta$ model: a fully-fledged chiral gauge theory once one adds an adjoint scalar to abelianize the $SU(N)$ gauge group.

Our main result, for the Abelianized $\psi\chi\eta$ model, is that the monopole triggers the formation of a CR condensate which preserves all symmetries apart from the anomalous one. The boundary conditions and scattering matrix also preserve the non-anomalous symmetries. The IN single particle states go into fractional, $1/4$ or $1/8$, OUT states, which we may call ``microtons''.%
The physical interpretation of the microtons remains an open problem.

The paper is organized as follows.
The first part of the paper serves as a more detailed introduction to the Callan-Rubakov effect. In particular, in Sec.~\ref{sec:dirac} we discuss the physics of fermions with an arbitrary electric charge on a Dirac monopole background and how the s-wave fermions scattering is described by a two-dimensional fermion theory, while in Sec.~\ref{sec:qed} we review the basic aspects of the CR effect in vectorial QED. In Sec.~\ref{sec:chiral} we give a generic introduction to the problem for chiral theories, define the space of locally vector-like theories, and present the Abelian quiver theory we aim to study.
Then we turn into the discussion of the CR effect in our chiral example, following the steps outlined before: in Sec.~\ref{sec:psichieta} we introduce the $\psi\chi\eta$ model, describe its IR $4$D physics, and check that the model admits heavy 't Hooft-Polyakov monopoles in its spectrum; in section Sec.~\ref{sec:rubakov}  we calculate the Callan-Rubakov condensate and in Sec.~\ref{sec:bosonization} discuss the bosonized picture in the  $\psi\chi\eta$ model and the scattering problem.
We conclude in Sec.~\ref{sec:conclude}.

\section{Charged fermions in a Dirac monopole background}
 \label{sec:dirac}

{ 

Let's begin discussing how to set up the problem in general. It is convenient to consider three theories
\beq{\rm UV}_{3+1} \longrightarrow {\rm IR}_{3+1} \longrightarrow {\rm IR}_{1+1}\;.\eeq
We are ultimately interested in the scattering of massless fermions of the $\rm IR_{3+1}$ theory on Dirac monopoles. The ${\rm UV}_{3+1}$ theory provides us a UV completion of the $\rm IR_{3+1}$ EFT, where the Dirac monopole is regularized, e.g. as a 't Hooft Polyakov monopole. Thus ${\rm UV}_{3+1} \longrightarrow {\rm IR}_{3+1}$ is characterized by an energy scale $\Lambda_{UV}$, which is roughly the size of the monopole. %
Once one considers ${\rm IR}_{3+1}$ in the presence of a Dirac monopole, it is convenient to decompose the low-energy fermions in modes with different angular momentum. Then, in the $\alpha_{EM} \ll 1$ limit, the waves with different angular momentum decouple: they interact only via the exchange of photons (suppressed by the $\alpha_{EM}$) or by the non-Abelian interactions inside the monopole core, process suppressed in the  $\omega \ll \Lambda_{UV}$ limit. 

In these two limits, one can solve the dynamics of the mode with higher angular momentum, without any information about ${\rm UV}_{3+1}$. On the contrary, the modes of minimal angular momentum interact with the monopole core, probing the UV physics. We will thus restrict to them, and we call $\rm IR_{1+1}$ the infrared theory that describes them. This theory lives on the $r>0$ half-line and the monopole core is seen as a boundary at $r=0$.
}

Finally, we will ignore the well-known infrared divergences that arise in any unbroken gauge theory. Accounting for them would require dressing our charges with soft clouds of massless particles, which is known to be problematic in the presence of massless gauge charges because a Poincarè covariant S-matrix is not permitted. First of all, we circumvent this problem by computing correlation functions at finite times. We will also take the point of view that the deep infrared dynamics takes over at very large times, so we will be discussing the \textquotedblleft hard\textquotedblright{} part of the scattering process. 

We will now start from the $\rm IR_{3+1}$ theory and study the fermionic modes on top of the monopole background. This section reviews the decomposition in angular momentum eigenmodes\footnote{See \cite{Kazama:1976fm}, or \cite{Shnir:2005vvi} for a textbook presentation.} and argues that modes with non-minimal angular momentum scatter trivially.
\medskip

	Let's take a Dirac monopole with magnetic charge $m\in \mathbbm{Z}$
		and a left-handed Weyl fermion $\psi$ with charge $q\in \mathbbm{Z}$. 
We define  the parameter
		\be
		\mu=\frac{1}{2}|mq|\in \frac{1}{2}\;\mathbbm{Z}\;.
		\ee
 The Dirac monopole background is invariant under rotations generated by the deformed angular momentum
		\be
		{\tilde L}=L - \frac{1}{2}mq \hat{r}
		\ee
Additionally, the Weyl equation on the monopole background is invariant under 
		\be
		J={\tilde L} + S \;.
		\ee
where $S$ is the angular momentum contributed by the spin of the fermion.

In appendix \ref{appendix:monopole_harmonics}, we explicitly build the spinor monopole spherical harmonics $\Omega_{\mu,j,m}^{(1)}$ and $\Omega_{\mu,j,m}^{(2)}$, which account for all non-minimal angular momentum modes, and $\Omega^{(3)}_{\mu,\mu-\frac{1}{2},m}$, which describes the lowest angular momentum mode. Notice that $\Omega^{(1)}_{\mu, j, m}$, $\Omega^{(2)}_{\mu, j, m}$ and $\Omega^{(3)}_{\mu, j, m}$ have a spinor index, not displayed here, which should not be confused with the $m$, which labels the $J_z$ eigenvalue.  On the monopole background, $A_i$, the spinor monopole harmonics greatly simplify the Weyl equation
\bea
	& \qquad \ \  \ \bar{\sigma}^i (\partial_i -i q A_i)\zeta(r)\Omega^{(1)}_{\mu, j, m}=-\left(-\frac{d}{dr} - \frac{1}{r}+ \frac{\tilde\ell}{r}\right)\zeta(r)\Omega^{(2)}_{\mu, j, m}&\nn \\
	& \qquad \ \  \  \bar{\sigma}^i (\partial_i -i q A_i)\zeta(r)\Omega^{(2)}_{\mu, j, m}=-\left(-\frac{d}{dr} - \frac{1}{r}- \frac{\tilde\ell}{r}\right)\zeta(r)\Omega^{(1)}_{\mu, j, m} &\nn \\
	& \bar{\sigma}^i (\partial_i -i q A_i)\zeta(r)\Omega^{(3)}_{\mu, j, m}=-\left(\frac{d}{dr} + \frac{1}{r}\right)\zeta(r)\Omega^{(3)}_{\mu, j, m} & \label{eq:importanteq}
	\eea
	where $\tilde\ell=\sqrt{(j+\frac{1}{2})^2-\mu^2}$.

	At this point, we decompose the Weyl fermion as
	\be
	\psi=\frac{1}{r}\Bigg\{ H_m(r, t) \Omega^{(3)}_{\mu, \mu-\frac{1}{2}, m} +  \sum_{j=\mu+\frac{1}{2}, ...} \sum^{j}_{m=-j} F_{j,m}(r, t) \Omega^{(1)}_{\mu, j, m} + iG_{j,m}(r, t) \Omega^{(2)}_{\mu, j, m} \Bigg\}\;.
	\ee	
	By plugging this decomposition into the Weyl equation
	\be
	\bar{\sigma}^\mu (\partial_\mu - i q A_\mu)\psi=0\;,
	\ee
	and using the identities (\ref{eq:importanteq}) we have, for the non-minimal angular momentum modes $j=\mu+\frac{1}{2}, \dots$, 
	\be
	\psi_{j, m}^{2{\rm D}} = \left(\begin{array}{c}
		\im F_m+G_m\\
		F_m+\im G_m
	\end{array}\right),\qquad
	\Bigg(\im \gamma^\mu\partial_{\mu}-\frac{\tilde{\ell}}{r}\Bigg)\psi^{2{\rm D}}_{j, m}=0\;, \label{eq:farmodes}
	\ee
	where the $2$D gamma matrices are defined as in (\ref{eq:gamma_matrices_2d_lorentz}).
    The minimal angular momentum $j=\mu-\frac{1}{2}$ mode of a positively charged left-handed fermion is purely in-coming
	\be
	\psi= \frac{H_m(t+r)}{r} \Omega^{(3)}_{\mu,\mu-\frac{1}{2}, m}
	\ee
	i.e. its dynamics is described by a purely left-moving 2D fermion $H_m$, obeying 
	\be(\partial_t - \partial_r)H_m=0\ee

	If the electric charge is negative, $q<0$,  the minimal angular momentum mode is
	\be
	\psi= \frac{H_m(t-r)}{r}  \; \left(\sigma^2 \Omega^{(3)}_{\mu, \mu-1/2, m}\right)^*\;.
	\ee
	This mode is purely out-going. 
A positively charged right-handed fermion is equivalent to a negatively charged left-handed fermion and is out-going too. Indeed plugging
\be
\psi=\frac{H_m(t-r)}{r} \Omega^{(3)}_{\mu, \mu-1/2, m}
\ee
into the Weyl equation and using eq.~(\ref{eq:importanteq}), we can check
\be
(\partial_0 + \sigma^i(\partial_i-i|q|A_i))\frac{H_m(t-r)}{r} \Omega^{(3)}_{\mu, \mu-1/2, m}=0\;. 
\ee

\paragraph{Convention:} We will denote the $4D$ spinors as $\psi_{L,R}$, while their s-wave reduction as $\psi_{l,r}$.

\subsection{Dynamics in the effective theory}
\label{bounce}
 
The assumption behind the study of the Callan-Rubakov effect is that one can focus on the \textquotedblleft s-wave\textquotedblright{} theory, which is effectively 2D, and throw away all the modes with $j>j_0=\mu-1/2$, while the dynamics of $j_0$ must be supplemented by a boundary condition.

We will show this to happen only for the lowest angular momentum mode, first more physically and then rigorously. Mathematically speaking, the problem is that the Hamiltonian might admit a non-unique self-adjoint extension, effectively defining different Hamiltonians and therefore different dynamics.

Physically, we would like to argue that higher angular modes can be dropped because they miss the monopole core, therefore experiencing the low energy dynamics only. To make this more concrete, we will now take the higher mode profiles and estimate what is the probability for the particle to sit inside the core. We can solve the equations of motion on the Dirac monopole background, obtaining (see eq.~(\ref{self-adjoin_app:psiL}) and discussion below)
\be
\psi_{l,r} = \Gamma(1-\l) e^{\mp i\omega r}(\mp i 2\omega r)^\l l_{-\l}^{2\l}(\pm 2ir\omega)
\ee
where we picked the only solution which is square integrable at the origin and normalized it at infinity. At $r\lesssim\frac{1}{\Lambda}=\text{monopole size}$, we expect the solution to deviate slightly, but to remain of the same order. The wavefunction inside the monopole core can then be estimated to be
\be
|\psi| \lesssim \Gamma(1-\l) \left(\frac{2\omega}{\Lambda}\right)^\l L_{-\l}^{2\l}(0)
\ee
which goes to zero for large $\Lambda$.
Even considering the infinity of higher $\l$ modes and their multiplicities, the product $\Gamma(1-\l)L_{-\l}^{2\l}(0)$ is always finite and actually vanishes as $\sim\l^{-\l}$, making any sum over $\l>0$ convergent. We conclude that the wavefunction inside the monopole core becomes arbitrarily small in the $\frac{\omega}{\Lambda}\rightarrow0$ limit even accounting for all $\l>0$.

We now sketch the more rigorous argument, referring to Appendix~\ref{appendix:self-adjoint_extension} for more detailed derivations. To study $j>j_0$, it is useful to work with  a single 2D Dirac fermion as defined in eq.~(\ref{eq:farmodes}), defining $\psi_{2{\rm D}} = (\psi_l,\psi_r)$.
Our problem is defined only for $r>0$, so imposing self-adjointness for the Hamiltonian is nontrivial due to the presence of the boundary. From the $2$D Dirac equation, ${\mathcal H}$ is   identified as
\be
i\partial_0\psi = -i\gamma^0\gamma^1\partial_r\psi+\frac{\tilde{\ell}}{r}\gamma^0\psi = {\mathcal  H}\psi
\ee
and the condition of self-adjointness is
\be
({\mathcal H}\psi,\phi) \equiv (\psi,{\mathcal H}\phi)
\ee
which leads to requiring
\be
\label{eq_questioning:self_adjoint_condition}
\int_0^\infty  \left(\psi_l^*\phi_l - \psi_r^*\phi_r\right)'\dd r = 0\;\ \implies \ 
\left( \psi_l^*\phi_l- \psi_r^*\phi_r\right)|_{r\to 0} = 0
\ee
where we assume that our energy eigenstates are actually smoothed out at large $r$ and go to zero sufficiently fast, so that they truly belong to the Hilbert space.

The problem can thus be set up as follows. Given the two solutions of the stationary Dirac equation, we can form the most general linear combination which is square integrable at the origin. Then we ask if eq.~(\ref{eq_questioning:self_adjoint_condition}) is trivially true or if it imposes additional conditions on the allowed linear combinations. We solve this problem in appendix \ref{appendix:self-adjoint_extension}, reporting here the main results.

For minimal angular momentum modes, $j=j_0$, the dynamics is determined up to a UV-dependent parameter $\theta$ which enters the boundary condition as
\begin{equation}
    \psi_l = e^{i\theta}\psi_r \label{eq:vectorialBC}
\end{equation}
Additionally, $\theta$ is an energy-independent parameter, so scale symmetry is preserved by the boundary condition.%

On the contrary, the boundary condition in the infrared theory is always uniquely specified when $j>j_0$ if we assume the unitarity of scattering on the monopole background.

Before proceeding, we want to stress that (\ref{eq:vectorialBC}) is the unique possibility only in quantum mechanics, once one identifies $\psi$ as the wave function, the vectorial current (fermion number)
\be
J^\mu=\left(\begin{array}{c}\bar{\psi}\gamma^0\psi \\ \bar{\psi}\gamma^1 \psi \end{array}\right)= \left(\begin{array}{c}\bar{\psi}_l\psi_l +\bar{\psi}_r\psi_r  \\ \bar{\psi}_l \psi_l -\bar{\psi}_r \psi_r \end{array} \right)
\ee
as the probability current, thus unitarity with its conservation on the boundary. 

However, in QFT, where one does not perform such identification and $U(1)_V$ is not a priori preferred, one can consider other boundary conditions, e.g.
\be
\psi_l = e^{i\theta} \bar{\psi}_r \label{eq:axialBC}
\ee
which conserves the axial current, but spoils the conservation of fermion number. We will see that the 't Hooft-Polyakov regularization of the monopole, if one does not integrate away the s-wave gauge fields, prefers (\ref{eq:vectorialBC}) over (\ref{eq:axialBC}), while, once one integrates them away and include their dynamics into the account in an effective boundary condition, in the presence of multiple fermion families, nor the axial nor the vectorial current of each fermion species is conserved by the effective boundary condition.

Higher spin modes play an important role in the mechanism discussed in \cite{Csaki:2022qtz}. We will comment on this later.

\section{The Callan-Rubakov effect in QED}
\label{sec:qed}
In this section, we will introduce the Callan-Rubakov effect in the simple setup of QED. We will see that the simplest case with one Dirac fermion is under control, while adding more fermions is problematic. We will also highlight the importance of the UV completion in determining the result of the scattering.

Let's take QED with $N_f$ massless Dirac fermions with charge $q$, and consider a single Dirac monopole with magnetic charge $m$. 
The solution to the Dirac equation for a charge $q$ Dirac fermion is therefore 
		\be
		\Psi=\sum_{m=-(\mu-1/2)}^{\mu-1/2}\frac{1}{r}\left(\begin{array}{c} f\big(t - {\rm sign}(qm) r\big)\Omega_{\mu, \mu-1/2, m}^{(3)} \\ g\big(t + {\rm sign}(qm) r\big)\Omega_{\mu, \mu-1/2, m}^{(3)}\end{array}\right) + \sum_{J>\mu-1/2} ...
		\label{psizeromodes}
		\ee
From now on we neglect the higher spin harmonics, and then focus only on $f(t\pm r)$ and $g(t \mp r)$r.
		For $qm>0$ ($qm<0$) the left-handed part of $\Psi$ is purely out-going (in-going), the right-handed part is purely in-going (out-going).
  
  For every 4D Dirac fermion, here for example with $qm=1$, what follows is the table of all left/right-handed particles/antiparticles of the 2D theory.
		\bea
			\begin{array}{c|c|c|c}
				& U(1)_{\rm e} &  4{\rm D\, chirality} & \text{direction}\\
				\hline
				\psi^+_L&1&\text{left}& \text{in}\\
				\psi^+_R&1&\text{right}& \text{out}\\
				\psi^-_L&-1&\text{left}& \text{out}\\
				\psi^-_R&-1&\text{right}& \text{in}\\
			\end{array}
   \label{table:qed_charges}
	\eea
	\smallskip

	The Dirac monopole has a singularity in the origin and has infinite mass, so we consider it as a background electromagnetic state, not a dynamical object.
 The $3+1$ abelian $U(1)$ theory is an effective model,  because it is IR free and   has a Landau pole, and also because the monopole background has a divergence in the core. 
	The dynamics of $\psi$ (or, equivalently, of $\Psi$) is not completely fixed by the equations of motion, because we need to fix the boundary condition at $r=0$. The correct boundary condition to impose depends, in principle, on the UV completion: as the fermion enter in the monopole core, it excites internal degrees of freedom, which react in a model dependent way. 
 From the $U(1)$ model and the effective $1+1$ theory   we can search for  possible consistent boundary conditions.

We begin by considering $N_f=1$. To determine the correct boundary condition, we can be guided by symmetries.
		In this simple case, there is only a single symmetry that acts non-trivially: $U(1)_{\rm e}$. One can fix the boundary condition to preserve $U(1)_{\rm e}$ by imposing
		\be
	     \psi_l|_{r=0}=e^{i\theta}\psi_r|_{r=0}\;.
		\ee
		With this boundary condition electric charge is preserved
		\be
		\psi^+_L  +  M \rightarrow \psi^+_R   +  M \;,
		\ee
		but chirality is not. 

 Notice that here we have implicitly made an assumption: there are no dyons degenerate to the monopole. If one drop this assumption one can also consider the alternative solution, 
  \be
		\psi^+_L  +  M \rightarrow \psi^-_L   +  D^{++} \;,
	\ee
 Only the UV theory can prescribe if there is such a dyonic state, thus, already here, we see a dependence on the putative UV phase.

  The situation is more interesting with higher $N_f$, and thus more symmetries.

\subsection{Embedding in $SU(2)$: the UV completion} \label{ssec:fermionembedding}

    To discuss the problem when there are more fermions, symmetry is not enough. To proceed further we must commit to a specific UV completion of QED. We will put the fermions on the 't Hooft-Polyakov monopole background and determine the low energy effective boundary condition.

	Let's take QED with $N_f$ massless fermions with minimal charge, $q=1$, and consider a single Dirac monopole with minimal magnetic charge. 
	This theory can be regularized by considering an $SU(2)$ broken to $U(1)$ by an adjoint scalar, with $N_f$ Weyl fermions in the fundamental representation. To avoid the Witten anomaly, this regularization makes sense only if $N_f=2k$.  
This theory enjoys an $SU(N_f)$ global symmetry, the $U(1)$ symmetry being broken to $\mathbbm Z_{N_f}$ by the ABJ anomaly.

Let us consider the $SU(2)$ resolution of the Dirac monopole into a  't Hooft-Polyakov monopole. 
We work in the hedgehog gauge, where the monopole has profile\footnote{The $h(r)$ and $k(r)$ profiles are fixed by two differential equations which are not important for our discussion.}
\bea
\phi  = v\;h(r)  \frac{r^a}{r} \sigma^a \qquad h(0)=0\;, \;\;h(r)\xrightarrow{r\rightarrow \infty} 1\;,\\
A_c=\epsilon_{abc}r^b\frac{1-k(r)}{r^2} T^a \qquad k(0)=1\;, \;\;k(r)\xrightarrow{r\rightarrow \infty} 0\;. \label{eq:tHooftPolyakov}
\eea
We now repeat the analysis of the previous section, now for the fermionic modes on top of the non-abelian monopole.
The background above is invariant under 
\be
J=L+S+T
\ee
where $T$ represents the $SU(2)$ gauge generators.

Let us consider a left-handed Weyl fermion $\eta_{i\alpha}$ in the fundamental\footnote{In this section, let us use a downstairs index for the fundamental rep of $SU(2)$. One can raise/lower the index with the $\epsilon$ tensor.} representation of $SU(2)$.
It is useful to define $\tilde\chi_{\alpha i}=\eta_{\alpha}^i$, and $\chi=\tilde \chi \epsilon$. Then the gauge transformation acts as
\be
\chi \rightarrow \tilde \chi U^T \epsilon = \tilde \chi \epsilon U^{-1}\;, \quad \phi \rightarrow U\phi U^{-1}\;.
\ee 
So one can classify the solutions of the Dirac equation in the background under the $J$ representations. We want to focus on the lowest possible representation of $J$, which for $T=\frac{1}{2},\;S=\frac{1}{2}$ irrep (fundamental fermions) is $J=0$. One can build a $J=0$ combining $L=1,\;S+T=1$ and $L=0,\;S+T=0$ 
\be
\eta_{i\alpha}=u(r,t)\epsilon_{i\alpha}+ v(r, t)\frac{r^a}{r} (\sigma^a)_{i\;}^{\;\beta} \epsilon_{\beta \alpha} %
\ee
or, using the matrix notation, 
\be
\chi=u(r, t)\delta+v(r, t)\frac{\sigma^a r^a}{r}\;.
\ee
By plugging the the $j=0$ modes into the Dirac equation one obtains 
\be
\begin{cases}
    \label{eq:fermion_nonabelian_monopole}
	\partial_t v+\partial_r u-\frac{k}{r} u=0\\
	\partial_t u+\partial_r v+\frac{k}{r} v=0
\end{cases} \qquad   \text{i.e.} \qquad 
\left(\gamma^i \partial_i -\gamma^5 \frac{k}{r}\right)\left(\begin{array}{c}-iu\\v\end{array}\right)=0\;,
\ee
in the non-chiral basis $\gamma^0=\sigma^3$, $ \gamma^1=-i\sigma^1 $, $ \gamma^5=\sigma^2$.

It is important here to note that, for $r\Lambda \lesssim 1$ ($\Lambda \gg \omega$), $k(r) \sim 1$  so $u(r)\sim r$, while  $v(r)\sim 1/r$. Thus we are forced to impose the boundary condition $v|_{r=0}=0$. 
With this caveat on the boundary condition, to study the IR physics we can drop the $\gamma^5 \frac{k}{r}$ mass term, as $k(r)\rightarrow 0$ far away from the monopole core. To make contact with (\ref{psizeromodes}) it is convenient to choose a chiral basis (\ref{eq:gamma_matrices_2d_lorentz}) for the Dirac equation
\be
\left(\begin{array}{c}\chi_l \\ \chi_r\end{array}\right)=\frac{1}{\sqrt{2}}\left(\begin{matrix}i & -1 \\ i & 1\end{matrix}\right)\left(\begin{array}{c}-iu\\v\end{array}\right)
\ee
so that 
\be
\gamma^i\partial_i \chi=0\;,\;\; \chi= \left(\begin{array}{c}\chi_l \\ \chi_r\end{array}\right) \rightarrow \partial_- \chi_l=0\;,\;\; \partial_+ \chi_r=0\;,
\ee 
supplemented by the boundary condition
\be \chi_l|_{r=0}=\chi_r|_{r=0}\;. \label{eq:mesoBC}\ee
Notice that, in a sense, this boundary condition is the unique interesting output of this analysis, and the fact that it confirms the expectation of single particle QM is not trivial. On the contrary, the fact that the fermion decomposes in an ingoing and outgoing massless 2D fermion is just a double-check on the decomposition in sec.\ref{sec:dirac}.

The $4$D solution  reads
\be
\chi=\frac{1}{\sqrt{8\pi}r} \chi_l(r+t)\left(1-\frac{r^a\sigma^a}{r}\right)+\frac{1}{\sqrt{8\pi}r} \chi_r(r-t)\left(1+\frac{r^a\sigma^a}{r}\right)
\ee
A radial $U(1)_{\rm e}$ gauge transformation $e^{i\alpha(r, t) \hat{r}\cdot\sigma}$ transforms each piece into itself, 
which is consistent with the charge assignment
\be
U(1): \chi_l \rightarrow e^{-i\alpha}\chi_l\;, \quad \chi_r \rightarrow e^{i\alpha} \chi_r\;, \label{eq:2DchiU1}
\ee
i.e. $\chi\rightarrow e^{i\alpha \gamma^5}\chi$.

Notice that the vectorial boundary condition, (\ref{eq:mesoBC}), does not respect the radial $U(1)_e$ transformation. Notice that, in the regular gauge, where the equation above is defined, the transformation above is singular at $r=0$. Indeed it respects the profile (\ref{eq:tHooftPolyakov}) only if $\alpha \ll 1$ for $r\sim \Lambda$. In other words, one should restrict to $\alpha \xrightarrow{r\to 0} 0$.

\subsection{Microscopic dynamics}

Let us consider a single Weyl fermion, $\eta^i_\alpha$, in the fundamental of $SU(2)$, using the same matrix notation as above. In the vacuum background $\langle\Phi\rangle=v\sigma^3$, the gauge group is higgsed, $SU(2)\rightarrow U(1)$,  and the left-handed Weyl fermion decouples into two left-handed Weyl fermions with $q=\pm 1$ (notice that $g_e=g_{SU(2)}/2$). 

If one performs this calculation in the fixed monopole background, one finds inconsistencies such as non-conservation of electric charge. The issue is that one cannot forget that the fermion carries electric charge which, if not the same in the IN and in the OUT state, must be transferred to the monopole.\footnote{Within the system in this section there are no degenerate dyons with different electric charge. This is because there are no Jackiw-Rebbi zeromodes. This is a feature of \textit{this} UV completion of QED, which might not hold for other models, so one should not rule out the possibility of a transfer of charge from the fermions to the dyons.}

To correctly take this into account one should consider electric and magnetic fluctuations on top of the monopole background, 
\be
A_0^a= \frac{r^a}{er}a_0(r, t) \quad A_m^a=\frac{\epsilon^{amn} r^n}{r^2}\left(1-k(r)\right) + \frac{r^n r^a}{er^2}a_1(r, t)\;. 
\ee
parameterized by the functions $a_0$ and $a_1$. One can plug this expression into the Lagrangian for the gauge field, and neglecting the core of the monopole ($M_{mon}\rightarrow \infty$) one obtains
\be
S_{\rm eff}=\frac{2\pi}{e^2}\int r^2 f^{ij}f_{ij}  + i \bar v \left(\gamma^i \partial_i + i \gamma^5 a_i\right)v\;. 
\ee
with the boundary conditions $v(r=0)=0$. This model can be solved. The trick is to change fermionic variables. We will discuss this technique in detail when we will apply it to our model of interest in Sec.\ref{sec:rubakov}.
  
Performing the condensate computation, one notices that there is a $N_f(=2k)$ fermionic condensate around the monopole, 
\be
\langle\underbrace{\epsilon^{a_1b_1}\epsilon^{\alpha_1, \beta_1}\psi^{A_1}_{\alpha_1, a_1}\psi^{A_2}_{\beta_1, b_1}...\epsilon^{a_k, b_k}\epsilon^{\alpha_k, \beta_k}\psi^{A_1}_{\alpha_k, a_k}\psi^{A_2}_{\beta_k, b_k} }_{N_f\;\text{fermions}} \epsilon^{A_1, ...A_{N_f}}
\rangle \propto \frac{1}{r^{3/2 N_f}}
\ee 
which keeps $SU(N_f)$ unbroken but breaks $U(1)_A \to \mathbbm{Z}_{N_f}$, realizing the ABJ anomaly.

\section{Chiral  theories}
\label{sec:chiral}
 
The examples of section \ref{sec:qed} all dealt with vectorial theories. The situation is even more fascinating for chiral theories because a fundamental mismatch between the charges of the ingoing and outgoing modes arises. 
It is worthwhile to give an example presented in Ref.\cite{Smith:2019jnh,Smith:2020nuf}. The model has a single $U(1)$ gauge group with five left-handed fermions with charges $1,5,-7,-8,9$. In 4D the theory has vanishing gravitational and gauge anomalies,  $[U(1)]=[U(1)]^3=0$, so it is a well-defined 4D chiral gauge theory. Once the theory is put on the Dirac monopole background, and only the lowest angular momentum modes are considered, one can produce the analog of table~\ref{table:qed_charges},

\beq
 \begin{array}{c|c|c|c|c|c|c|c}
 	&U(1)_{\rm e}&U(1)_{G1} &U(1)_{G2}&U(1)_{G3}&U(1)_A& SU(2)_{rot} & {\rm direction}  \\ \hline
 	\psi_{+1}    &+1        &+25      & 0       & 0    &+1           & 1 & \text{in}    \\
 	\psi_{+5}    &+5        &-1       & +49     & 0    &+25& 5 & \text{in}   \\
 	\psi_{-7}    &-7        & 0       & -25     & +64  &-49 & 7 &\text{out}   \\
 	\psi_{-8}    &-8        & 0       & 0       & -49  &-32 & 8 &\text{out}    \\
 	\psi_{+9}    &+9        &0        & 0       & 0    &-109 & 9 & \text{in}   
 \end{array} 
 \label{chiral1a}
 \eeq

Alongside the direction in which each mode travels in the 2d theory, there are its $SU(2)_J$ multiplicity, its charges under the ABJ-anomaly free symmetries of the 4D theory, $U(1)_{G1}$, $U(1)_{G2}$ and $U(1)_{G3}$, and under the anomalous one\footnote{Recentrly it has been understood that also the ABJ-anomalous symmetry should be better understood as an accidental non-invertible symmetry\cite{Cordova:2022ieu}. The magnetic monopole breaks this accidental symmetry.}, $U(1)_A$ (clearly the particular value of these charges depends on the basis of the abelian symmetries one prefers). 
As for the standard vectorial case, with the outgoing fermions alone, it is impossible to construct a multiparticle state that conserves all the charges of a given, incoming fermion. 

Before trying to solve the hard problem of constructing the correct scattering states a preliminary step is to determine which symmetries are preserved by the monopole boundary condition.
The simplest guess, analog to Callan-Rubakov's result in the vectorial case, is that $U(1)_A$ only gets explicitly broken. However, this result would be problematic: it has been noticed that the symmetries participating in a 2-group structure with the magnetic 1-form symmetry, $U(1)^{(1)}_m$ (see \cite{Cordova:2018cvg} for a gentle introduction on the subject) are broken by the monopole\footnote{This is because, by reducing to the 2D EFT, the 2-group structure implies a 2D 't Hooft anomaly for the symmetry, which is incompatible with a symmetry preserving boundary condition \cite{Jensen:2017eof,Thorngren:2020yht}.} \cite{vanBeest:2023dbu}.  Both $U(1)_{G1}$, $U(1)_{G2}$ and $U(1)_{G3}$ form a 2-group with $U(1)^{(1)}_m$, thus all of them are broken to a subgroup. Unfortunately, a UV completion for this model is unknown, thus we cannot proceed as we did in section~\ref{sec:qed}, and verify what happens. 

Given a generic IR $U(1)$ theory it is hard, in general, to provide a UV completion. Let's discuss a partial obstruction. Consider a theory where the UV gauge group is $SU(N)$, and suppose that it completely abelianizes to its Cartan in the IR, leading to an $U(1)^{N-1}$ theory. Now take a basis where $U(1)_i$ is generated by 
\be
\text{diag} \overbrace{ (\underbrace{0, ..., 0}_{i-1}, 1, -1, 0, ..., 0)}^{N}.
\ee 
This is also the Cartan generator of $SU(2)_i$, an $SU(2)$ that minimally embeds into $SU(N)$. Decomposing any representation of $SU(N)$ in the representation of $SU(2)_i$ first, and then under $U(1)_i$, one obtains that for any particle with charge $q$, there is another one with charge $-q$: the theory is \textit{locally vector-like}, i.e. there exists a basis of the Cartan, such that, considering the charges under each Cartan at a time, the spectrum is vector-like.

Notice that a locally vector-like theory can be chiral. Indeed, in the rest of the paper, we will analyze a chiral, but locally vector-like, theory that descends from a (chiral) $SU(N)$ theory by abelianization.

The situation does not change if one considers a more general UV completion, where one starts from a semisimple gauge group and abelianizes it to its Cartan. If one embeds a 't Hooft Polyakov monopole minimally into the UV gauge group, the resulting spectrum is always vector-like under the $U(1)$ in which the resulting Dirac monopole sits.

Now let's limit ourselves to considering $SU(N)$ factors. A non-minimal monopole can be obtained by embedding the $SU(2)$ generators of the 't Hooft-Polyakov in the $SU(N)$ algebra non-minimally. For example, to obtain a maximal embedding one should realize $SU(2)$ as its $N$-dimensional irrep. Still the leftover $U(1)$ that carries the magnetic field sits inside an $SU(2)$ group, thus the argument above continues to apply.
Moreover, in the case of non-minimal embeddings, there are light modes of the gauge fields trapped in the monopole core. For an analysis of these modes, and how they contribute to the CR problem, see \cite{Brennan:2021ewu}. 

One can evade this obstruction by considering further breaking of the gauge group, e.g. 
\be
SU(N) \rightarrow U(1)^{N-1} \rightarrow U(1)\;.
\ee
In this case, the spectrum is not (in general) vectorlike for the resulting $U(1)$.
However, this scenario does not simplify our problem, as the resulting magnetic monopole might have an internal structure.
Because of these facts, we will take a chiral but locally vector-like theory as our main example.

The first paper to discuss the CR effect in a chiral theory, starting with a well-defined UV completion, was the Callan one  
 \cite{Callan:1983tm} where the GUT monopole $SU(5) \to SU(3) \times SU(2) \times U(1)$ was considered.  However, the chiral nature of this model is intrinsically related to the nonabelian nature of the standard model gauge group. If we force a breaking pattern with complete abelianization, or even if we consider the theory below the Higgs scale, we have a vectorial model. 

 Here we aim to find a simple chiral $U(1)$ theory, without leftover non-abelian factors, which is chiral but locally vector-like.  One of the simplest example is built with two $U(1)$'s and a set of  Weyl left-handed fermion with charges $8\cdot(1,0)$,$8\cdot(-1,1)$, $8\cdot(-1,0)$,$1\cdot(-2,0)$,$1\cdot(2,-2)$, $1\cdot(2,0)$. This is locally vectorial, as already manifested in this base for the $U(1)$'s. Being locally vectorial all anomalies $[U(1)_1]$, $[U(1)_2]$ and  $[U(1)_1]^3$,  $[U(1)_2]^3$ are sure to be vanishing. Moreover, $[U(1)_1][U(1)_2]^2$  and  $[U(1)_1]^2[U(1)_2]$ vanishes too due to the number of fermions we have chosen\footnote{A more general class of theories of this kind is a $U(1)_1 \times U(1)_2$ gauge theory with fermions $q^3\cdot(p,0)$, $q^3\cdot(-p,p)$,  $q^3\cdot(-p,0)$, $p^3\cdot(-q,0)$, $p^3\cdot(q,-q)$, $p^3\cdot(q,0)$, with $p$ and $q$ positive integers. The simple case corresponds to $p=1$, $q=2$.}. 
\beq
\small
\begin{tabular}{c|c|c|c|c|c|c|c}
&$U(1)_1$  &$U(1)_2$   &$U(1)_{G1}$&$U(1)_{A1}$&$U(1)_{A2}$& $U(1)_{A3}$  &$SU(8)$  \\
\hline
$\eta_{1} $ &$+1 $&$0$ &$1 $&$ 0$& $ 0$ & $+ 4$ &8   \\
$\eta_{2} $ &$-1 $&$+1$ &$ 1 $&$ +1$& $ 0$ &$ +4$ &8   \\
$\eta_{3} $ &$0 $&$-1$ &$ 1 $&$0$&$ +1$ & $+ 4$ &8    \\
$\psi_{1} $ &$+2 $&$0$ &$ -2 $&$ 0$& $ 0$ &$+ 1$  &1    \\
$\psi_{2} $ &$-2 $&$+2$ &$ -2 $&$ -4$&  $ 0$&  $+ 1$&1    \\
$\psi_{3} $ &$0 $&$-2$ &$ -2 $&$0$&   $ -4$& $ +1$&1    \\
\end{tabular}
\label{quiv1}
\eeq 

Notice that an interesting feature of this model is that also the 2-group structure constants of $U(1)_{G1}$, given respectively by the $U(1)_1 U(1)_{G1}^2$ and $U(1)_2 U(1)_{G1}^2$ anomaly coefficients, vanishes\footnote{This fact is not a coincidence: as the UV completion is a theory with multiple fermion species, there is an ABJ anomaly free $U(1)$. Upon abelianization, the fact that there is no 2-group structure constant for the global symmetries already present in the UV model is simply a consequence of the absence of $SU(N) U(1)^2$ anomalies.}: $U(1)_{G1}$ can be unbroken.

In the following, we will study this model, first by providing a UV completion and then working out the dynamics of the relevant degrees of freedom to recover the Callan-Rubakov effect.

\section{Monopoles and fermions in the $\psi\chi\eta$ model} \label{sec:psichieta}

		The $\psi\chi\eta$ model  was  studied earlier  in \cite{Goity:1985tf,Eichten:1985fs,Armoni:2012xa}   and more recently in   \cite{Bolognesi:2017pek,Bolognesi:2022beq,Sheu:2022odl}.
		It  is  an $SU(N)$ gauge theory  with left-handed fermion matter fields  
		\be   \psi^{\{ij\}}\;, \qquad  \chi_{[ij]}\;, \qquad    \eta_i^A\;,\qquad  A=1,2,\ldots 8\;,  
		\ee
		a symmetric tensor,  an  anti-antisymmetric tensor and eight  anti-fundamental multiplets  of $SU(N)$,  or
		\be     \yng(2) \oplus  {\bar  {\yng(1,1)}}  \oplus   8  \times  {\bar  {\yng(1)}}\;.  \label{fermions2}
		\ee
				The global symmetry group is 
		\be   G_{\rm f}= \frac{U(1)_{\psi\chi} \times  {\tilde U}(1)\times  SU(8)}{   {\mathbbm Z}_N    \times  {\mathbbm Z}_{8/ N^*} } \;.  \label{symmetry}
		\ee
		where $N^* = {\rm gcd}(N+2,N-2)$. The  $\tilde U(1)$ acts as 
		\be
		\psi^{ij} \rightarrow e^{2i\alpha}\psi^{ij} \quad \chi_{ij} \rightarrow e^{-2i\alpha}\chi_{ij} \quad \eta^A_i \rightarrow e^{-i\alpha}\eta^A_i 
		\ee
		and  $U(1)_{\psi\chi} $ acts as 
		\be
		\psi^{ij} \rightarrow e^{i\frac{N-2}{N^*}\beta}\psi^{ij} \quad \chi_{ij} \rightarrow e^{-i\frac{N+2}{N^*}\beta}\chi_{ij} \quad \eta^A_i \rightarrow \eta^A_i 
		\ee
		while $SU(8)$ acts on the $\eta$ only.
				The division by  $ {\mathbbm Z}_N  $   is due to the fact that the color $ {\mathbbm Z}_N  $ center is shared by a subgroup of the flavor $U(1)$ groups. 

		So far we have not introduced any Higgs. The IR fate of this theory is not known for certain, but we know that the theory, because of 't Hooft anomaly matching (one cannot to write color simple color singlet states that matches all the 't Hooft symmetries) probably breaks some of the symmetries, gauge or/and global.
		A rather simple possibility advocated in \cite{Bolognesi:2017pek,Bolognesi:2022beq,Sheu:2022odl}  is dynamical abelianization, where it develops a bilinear condensate, 
		\be
		\langle  \psi^{ij}\chi_{ik} \rangle \sim {\rm diag}(a_1, a_2, ...., a_n)\;,  \qquad a_i \simeq \Lambda^3\;,
		\ee
		which breaks both the global and the   symmetry groups
		\be
		SU(N)_{\rm c} \rightarrow U(1)^{N-1}\;, \qquad U(1)_{\psi\chi} \rightarrow \mathbbm Z_{4/N^*}\;.
		\ee
		This process is dynamical and (if happens) it happens at strong coupling.

		 To have Abelianization while the theory is weakly coupled, it is useful to deform the $\psi\chi\eta$ model by adding an $SU(N)$ adjoint scalar (complex or real) $\Phi^a_b$, which leads to the theory we will consider. One then can choose the potential such that the theory abelianizes at high energy,
		\be
		\Phi={\rm diag}(a_1, a_2, ...., a_n) \qquad 
  \qquad 
  |a_i - a_j|_{i \neq j} \gg  \Lambda ,
		\ee 
		where $\Lambda$ is the strong coupling scale of the theory. This vev breaks $SU(N)\rightarrow U(1)^{N-1}$.

				It also possible to add a Yukawa coupling
		\be
		y\Phi^a_b \psi^{\{b, c\}}\chi_{[c, a]}=\frac{y}{2}\Phi^a_b\left(\psi^{\{b, c\}}\chi_{[c, a]}-\psi^{\{b, c\}}\chi_{[a, c]}\right)\;.
  \label{yc}
		\ee
		As the theory abelianizes the Yukawa term gaps together $\psi^{ij}$ for $i \neq j$ and all $\chi_{ij}$. In particular,
		\be
		y\Phi^a_b \psi^{\{b, c\}}\chi_{[c, a]}=\frac{y}{2}\left(a_a \psi^{\{a, c\}}\chi_{[c, a]}-\psi^{\{a, c\}}\chi_{[a, c]}\right)=\frac{y}{2}\left(a_i-a_j\right)\psi^{\{ij\}}\chi_{[i, j]}\;.
		\ee	
		The other fermions, $\psi^{ii}$ and $\eta^A_i$, $i, j=1,.., N$, $A=1, ..., 8$, remain massless.
		
If we  
choose $\Phi$ complex the Yukawa coupling breaks the $\Phi$ number symmetry and $U(1)_{\psi\chi}$ down to the diagonal subgroup, $U(1)_{\psi\chi}'$
			\be
			\psi^{ij} \rightarrow e^{i\frac{N-2}{N^*}\beta}\psi^{ij} \quad \chi_{ij} \rightarrow e^{-i\frac{N+2}{N^*}\beta}\chi_{ij} \quad \eta^A_i \rightarrow \eta^A_i \quad \Phi \rightarrow e^{-i\frac{4}{N^*}\beta}\Phi \;.
			\ee
			  In particular, as $\Phi$ take the vev, $U(1)_{\psi\chi}$ is spontaneously broken, and the overall phase of $\Phi$ corresponds to the NGB of the theory. 
 If we choose $\Phi$ real there is not any extra $\Phi$ number $U(1)$ symmetry. If we write anyhow the Yukawa coupling (\ref{yc}), 
			we break $U(1)_{\psi\chi}\rightarrow \mathbbm Z_{4/N^*}$ spontaneously. The situation is simpler, as there is no NGB.
	In the following, we focus for simplicity on the first one $\Phi$ real (or complex + $U(1)_{\psi\chi}$ breaking potential).

        As one integrates out all the massive fermions, namely the $\psi^{ij}$ with $i\neq j$ and the $\chi_{ij}$, and the massive vector bosons, what remains at low energy is chiral $U(1)^{N-1}$ quiver theory, a generalization of the one discussed before in (\ref{quiv1}). A particularly convenient choice of a basis for the gauge group is
        $U(1)^{N-1}=U(1)_1 \times ... \times U(1)_i \times ...\times U(1)_{N-1}$, where $U(1)_i$ is generated by 
        \be
		Q={\rm diag}(\underbrace{0, ..., 0}_{i-1}, 1, -1, \underbrace{0, ..., 0}_{N-i-1}) \ .
  \label{Qgenerator}
		\ee
With this choice of basis for $U(1)^{N-1}$ the kinetic matrix is not diagonal.  With this choice the leftover fermions $\psi^{ii}$ and $\eta^A_i$ are charged only under $U(1)_{i-1}$ and $U(1)_i$, so they organizes themselves in a quiver diagram
        \bea
	\qquad	
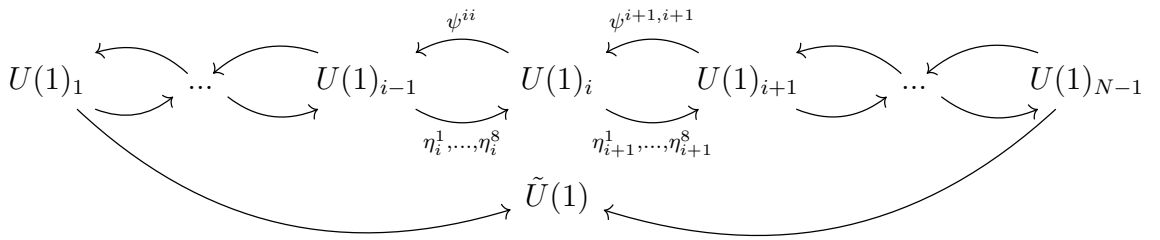
\begin{tikzcd}%
			 U(1)_1 \arrow[r, bend right] \arrow[rrrd, bend right]  &  |[draw=none]|... \arrow[l, bend right] \arrow[r,  bend right] & U(1)_{i-1} \arrow[r, "{\eta^1_i, ..., \eta^8_i}"', bend right] \arrow[l,  bend right] & U(1)_i \arrow[l, "\psi^{ii}"', bend right] \arrow[r, "{\eta^1_{i+1}, ..., \eta^8_{i+1}}"', bend right] & U(1)_{i+1} \arrow[l, "{\large \psi^{i+1, i+1}}"', bend right] \arrow[r,  bend right] &  |[draw=none]|... \arrow[r, bend right] \arrow[l, bend right] & U(1)_{N-1} \arrow[l, bend right] \arrow[llld, bend left]  \\
			&&&|[rectangle]| {\tilde U(1)}&&&
		\end{tikzcd}  \nn
        \eea

		We want to study the monopoles of the theory and how they couple to fermions.
To be concrete, let us take the $U(1)_i$ generator inside $SU(N)$ as (\ref{Qgenerator})		
		and let us imagine it embedded in a 't Hooft-Polyakov monopole for the $SU(2)$ that acts only on the subspace with $SU(N)$ index $j=i, i+1$.

		\subsection{JR zeromodes}
		
		Before we analyze the zero-modes of the $\psi\chi\eta$ model, let us review how the story goes for a Dirac fermions which transforms in the fundamental of $SU(2)$ and gets a mass both from a Yukawa coupling ($SU(2)\rightarrow U(1)$) and from a Dirac $SU(2)$ invariant mass term, within the background of a 't Hooft-Polyakov monopole.
		 
		Let us take a theory of $N$ $SU(2)$ fundamental Dirac fermions $\Psi^{a,A}$, being $a=1, 2$ the $SU(2)$ index and $A=1, ..., N$ the flavor index, in the background of a 't Hooft-Polyakov monopole. We can add a Dirac and a Yukawa mass terms, 
		\be
		M_{AB} \bar{\Psi}_{a,A}\Psi^{a, A} + y_{AB} \Phi^a_b \bar{\Psi}_{a, A}\Psi^{b, B}. 
		\ee  
		Upon breaking $SU(2)\rightarrow U(1)$, $\Psi ={\rm diag}(a, -a)$, the $SU(2)$ Dirac fermion splits in a Dirac fermion with charge $+\frac{1}{2}$ under $U(1)$ and a fermion with charge $-\frac{1}{2}$ under $U(1)$. Their masses are $M_A \pm y_A a$.  
		
	 	In this setting Harvey \cite{Harvey:1983tp} checked that in the $SU(2)$ 't Hooft-Polyakov background:
	 	\begin{itemize}
	 		\item If $M_A>y_A a$ there are no normalizable zeromodes
	 		\item If $M_A<y_A a$ there is a single real zeromode. Let us call its creation operator $a_A$, $a_A^2=0$.
	 	\end{itemize}
		If among the $N$ Dirac fermions $N^*$ satisfy $M_A<y_A a$, then there are $N^*$ zeromodes. From the quantization of them on obtains a Clifford algebra (the zeromodes creation operator satisfy $\{a^A, a^B\}=\delta^{AB}$, i.e. they are $\Gamma$ matrices).
		
		Let us apply the same idea on the $\psi\chi\eta$ model. As we are interested only on fermions coupled with $U(1)_i$, we care only about 
		\be
		\psi^{aJ}, \psi^{ab}, \chi_{a,J}, \chi_{a,b}, \eta^A_a \quad \text{where}\quad a,b=i, i+1\;,\quad J=1, ..., \check{i}, \check{i+1}, ..., N, \quad A=1, ..., 8\;. 
		\ee
	    Through the Yukawa coupling, $\psi^{aJ}$ is gapped together with $\chi_{aJ}$. In particular one can build $N-2$ Dirac fermion
	    \be
	    \Psi^a_J=\left(\begin{array}{c}\psi^{aJ}_\xi \\ (i\sigma^2)_{\dot\xi\beta}\bar{\chi}^{aJ}_\beta\end{array}\right) \quad \text{for} \quad J=1, ..., \check{i}, \check{i+1}, ..., N\;.
	    \ee
	    Then the Yukawa coupling, in the vacuum gives a mass term
	    \be
	    \frac{y}{2}\left(\bar\Psi^{i,J},\bar \Psi^{i+1,J} \right)\left(\begin{matrix}a_J-a_i & 0\\ 0 & a_J-a_{i+1}\end{matrix}\right)\left(\begin{array}{c}\Psi^{i,J},\\ \Psi^{i+1,J}\end{array}\right)
	    \ee
	    Comparing with the general result, one obtains that the fermion generate the zeromode only if 
	    \be |a_J-\frac{a_i+a_{i+1}}{2}|<|a_{i+1}-a_{i}| \label{eq:dis_masterJR}\ee
	    
	    Up to a gauge transformation, we can always order the eigenvalues of $\Phi$ such that $a_1<a_2<...<a_N$. Doing so, it is particularly simple to study the monopole embedded into the $U(1)_i$ such that $a_{i+1}-a_i$ is the smallest difference between the $a_j$s. In this case, the condition (\ref{eq:dis_masterJR}) is never fulfilled, so there are no Jackiw-Rebbi zeromodes linked to $\psi^{aJ}$, $\chi_{aJ}$ fermions.
        \medskip

        To completely rule out the existence of JR zero-modes, all that is left to do is check the non-existence of time independent solutions that remain bounded at the origin and decay at infinity for the remaining 6 fermions, which obey the matrix equation
        \begin{equation}
            \mathbf R\deriv{}{r}\zeta + \mathbf B(r)\zeta + \mathbf M(r)\bar\zeta = 0
        \end{equation}
        where $R^*=-R,B^*=-B,M^*=M$.
        Separating modes in real and imaginary parts, we get
        \begin{equation}
            i\begin{pmatrix}
                \mathbf R&\\
                &\mathbf R
            \end{pmatrix}\deriv{}{r}\begin{pmatrix}
                \zeta_r\\ \zeta_i
            \end{pmatrix} = 
            \begin{pmatrix}
                -i\mathbf B(r)&-\mathbf M(r)\\
                -\mathbf M(r)&-i\mathbf B(r)
            \end{pmatrix}
            \begin{pmatrix}
                \zeta_r\\ \zeta_i
            \end{pmatrix}
        \end{equation}
        Inverting the matrix ${\small \begin{pmatrix}
            \mathbf R&\\
            &\mathbf R
        \end{pmatrix}}$ and changing the basis $\zeta\rightarrow\zeta'$ by a permutation, we can bring the resulting matrix to block diagonal form. There are four blocks, each of which is a $3\times 3$ matrix $\mathbf A(r)$ up to a sign. The result is
        \begin{equation}
            \deriv{}{r} \zeta' = \begin{pmatrix}
                \mathbf A&&&\\
                &-\mathbf A&&\\
                &&-\mathbf A&\\
                &&&\mathbf A
            \end{pmatrix}\zeta',\quad
            \mathbf A(r) = \begin{pmatrix}
                \frac{1}{r}&-\frac{y h(r)}{\sqrt{3}}&-\sqrt{\frac{2}{3}}y h(r)\\
                -\frac{y h(r)}{\sqrt{3}}&\frac{-1+4k(r)}{3r}&\frac{\sqrt{2}(k(r)-1)}{3r}\\
                -\sqrt{\frac{2}{3}}y h(r)&\frac{\sqrt{2}(k(r)-1)}{3r}&-\frac{2(2k(r)+1)}{3r}
            \end{pmatrix}
        \end{equation}
        where we recall that the monopole profile functions $k(r),h(r)$ have the limits
        \begin{equation}
            \begin{tabular}{cc}
                $k(0)=1$, & $k(\infty)=0$ \\
                $h(0)=0$, & $h(\infty)=1$ \\
            \end{tabular}
        \end{equation}
        and we put the size of the monopole to $r=1$. We assume monotonicity of $k(r)$ and $h(r)$. The equations to solve are
        \begin{equation}
        \label{eq:JRmodes_eq}
            \deriv{}{r}\xi = \pm \mathbf A(r)\xi
        \end{equation}
        The eigenvalues of $\mathbf{A}(r)$ are $\frac{1}{r},\frac{1}{r},-\frac{2}{r}$ close to the origin, while they are $-y,0,y$ at $r\rightarrow\infty$. At finite but large distances, an analysis of the characteristic polynomial shows that the zero eigenvalue is actually positive and asymptotic to $\frac{2k(r)^2}{r^3}$. We will now proceed to show that the regular solutions at small distances cannot match the solutions regular at large distances, under the assumption of small Yukawa coupling, so that the $yh(r)$ term can be neglected up to a large scale $R_{\rm IR}$ much larger than the monopole scale: $R_{\rm IR} y h(R_{\rm IR})\sim 1$.
        
        We begin by focusing on the plus sign in eq.~(\ref{eq:JRmodes_eq}). A bounded solution at infinity should align with the eigenvector of eigenvalue $-1$, which is $(\sqrt{3},1,\sqrt{2})$. On the other hand, neglecting $yh(r)$ at $r\ll R_{\rm IR}$ allows us to solve for $\xi_1$, getting $\xi_1(r)=r$ up to an overall scaling factor which we set to 1. We trust this solution until $r\lesssim R_{\rm IR}$, where the decaying exponential takes over and $\xi_1\simeq R_{\rm IR}$. To align with the eigenvector $(\sqrt{3},1,\sqrt{2})$, we must have $\xi_2(R_{\rm IR})\sim\xi_3(R_{\rm IR})\sim R_{\rm IR}$ up to $\mathcal{O}(1)$ factors, but this contradicts the observation that, outside the monopole core (i.e. where $k(r)$ is small), $\deriv{\xi_{2,3}}{r}<0$ if $\xi_{1,2,3}>0$. Thus $\xi_{2,3}$ would have to grow up to $R_{\rm IR}$ but their derivative quickly becomes negative, contradicting the existence of solutions bounded everywhere.

        We now take the minus sign in eq.~(\ref{eq:JRmodes_eq}). This time solutions bounded at the origin see $\xi_3\simeq r^2$, where again we fixed an overall factor to $1$. $\xi_{1,2}$ begin with null slope, but are then driven to positive values by $\xi_3>0$. We now want to argue that if $\xi_{1,2,3}>0$, then they remain always positive. For $\xi_3$ this can be read off directly from $\xi_3'$, which is positive under our assumptions. For $\xi_1'$ we can rewrite the equation as $\frac{1}{r}\left(r\xi_1\right)'=\frac{yh(r)}{\sqrt{3}}\left(\xi_2+\sqrt{2}\xi_3\right)>0$, reaching the same conclusion. Finally, $\xi_2'$ is driven up by the terms proportional to $\xi_1$ and $\xi_3$, while the sign of $\frac{1-4k(r)}{3r}\xi_2$ depends on $k(r)$. However we can determine that $\xi_2$ will never cross zero by observing that this last term becomes negligible when $\xi_2\gtrsim0$, and the remaining positive terms keep $\xi_2'>0$. We conclude the argument by observing that, at large $r$, the eigenvectors of $-\mathbf{A}$ with negative eigenvalues are $(\sqrt{3},-1,-\sqrt{2})$ and $(0,\sqrt{2},-1)$ ($0$ is replaced by $\frac{\sqrt{6}k(r)}{y r}$ at large $r$), which cannot be combined to give a vector with only positive entries. This implies a mixing with the exponentially growing solution at infinity, precluding the existence of regular solutions.
	    
	    \subsection{CR zeromodes}
	    
		As $\psi^{ij}$ and $\chi_{ij}$, $i\neq j$, get a mass together, the remaining massless fermions are $\psi^{ii}$ and $\eta_i^A$. As we are looking at a monopole for $U(1)_i$, we are interested only at the fermions charged under $U(1)_i$: all the rest feel a centrifugal potential barrier, and, at the lowest order in the coupling constant, scatter without noticing the monopole. The interesting fermions are thus
		\bea
			\begin{array}{c|c|c|c|c|c|c|c|c} 
				\phantom{\tilde{\tilde{U}}}& U(1)_i & U(1)' & U(1)_{i-1} & U(1)_{i+1} & \tilde U(1) & SU(8) & SU(2)_{J} & 2d-\chi\\
				\hline
				\psi^{ii} & 2 & 2 & -2 & 0 & 2 & (\cdot) & \textbf{2} & \text{left } \\  
				\psi^{i+1, i+1} & -2 & 2 & 0 & 2 & 2 & (\cdot) & \textbf{2} & \text{right } \\
				\eta^A_i & -1 & -1 & 1 & 0 & -1 & \yng(1) & \textbf{1} & \text{right } \\
				\eta^A_{i+1} & 1 & -1& 0 & -1 & -1 & \yng(1) & \textbf{1} & \text{left} \\ 
			\end{array} \nn\\
\eea

        Here we have collected the first which these fermions are charged, $U(1)_i$ and $U(1)'$. In particular, it is convenient to choose as generators of
        \be
        U(1)_i:\; \text{diag}(\underbrace{0, ..., 0}_{i-1}, 1, -1, \underbrace{0, ..., 0}_{N-i-1}) \quad U(1)':\; \text{diag}(\underbrace{0, ..., 0}_{i-2},-1, 1, 1,-1, \underbrace{0, ..., 0}_{N-i-2})\;.
        \ee
        With this definition $U(1)_i$ is orthogonal to $U(1)'$. Comparing the coupling constant of $SU(N)$, $g$, with the ones of $U(1)_i$, $e_i$ and $e'$ of $U(1)'$, we have at the lowest order
        \be
        e_i=\frac{1}{2}g(\mu_b) \qquad e'=\frac{1}{2\sqrt{2}}g(\mu_b)
        \ee
        where $\mu_b$ is the approximate scale of the symmetry breaking.

        It is then useful to pack together the 2D fermions to obtain Dirac ones, in particular
        \be
        \psi^m_{2{\rm D}}= \left(\begin{array}{c}
             \psi^{ii, m} \\
              \psi^{i+1, i+1,m}
        \end{array}\right) \quad \eta^A_{2{\rm D}}= \left(\begin{array}{c}
             \eta^A_{i+1} \\
              \eta^A_{i}
        \end{array}\right)\;. 
        \label{dirac2dpsieta}
        \ee
        Notice that the upper component of the 2D Dirac fermions have the same charge under $U(1)_i$, because of the interlock between charge and 2D chirality (the upper components are always left-movers, with our choice of 2D $\gamma$ matrices).

        Summarizing, the 2D fermions have the following charges under $U(1)_i \sim U(1)_A$ and $U(1)' \sim U(1)_V$
           \beq
                     \begin{array}{c|c|c|c|c}
                 & U(1)_A & U(1)_V & SU(2)_J & SU(8)_\eta\\
                 \hline
                \psi_{2{\rm D}} & -2 & 2 & j=\frac{1}{2} &  \yng(1)  \!\!\!\!\!\!\!\!\phantom{\bar{\yng(1)}}\\ 
                \eta_{2{\rm D}} & -1 & -1 & j=0 & (\cdot)\\ 
            \end{array}  
            \label{chargespsieta2d}
        \eeq
        
        As we see, within this subset of fermions, there are some redundant symmetries: a linear combination of $U(1)_{i+1}$ and $U(1)_{i-1}$ acts exactly as $U(1)_i$, another one acts as $\tilde U(1)$. Clearly, this does not extend to the whole theory, there are other fermions, $\psi^{i+2, i+2}$, that distinguish between these symmetries.
		
As we reviewed in general, the first question to ask is if it is possible to have boundary conditions on the gauge theory that respect all the symmetries of the theory. In particular, we have to ask that

\be
J^I_x|_{r=0}=0 
\ee
for $I$ which labels the symmetries. If one restricts itself to consider only the $U(1)$ symmetries (i.e. from $SU(8)$ and $SO(3)_{
m rot}$ takes only the Cartan subgroups) the boundary conditions can be simply stated in bosonization language.

\subsection{Gauge fields}
 
These $U(1)$'s are crucially different in the way they are embedded into $SU(N)$. Indeed for their fluctuations $A$ we have
\begin{align}
	A_0 = a_0(r) \hat{r}\cdot\tau,\quad A_i = a_1(r) \hat{r}_i (\hat{r}\cdot\tau)\, , \qquad \text{for }\  U(1)_A \nn \\
	\tilde A_0 = \tilde a_0(r) \lambda,\quad \tilde A_i = \tilde a_1(r) \hat{r}_i \lambda\, , \qquad \text{for }\ U(1)_V
\end{align}
where $\tau$ is the generator of $U(1)_A$, which embeds into $SU(N)$ as $\hat{r}\cdot\tau$ at position $i,i+1$, while $\lambda$ is the generator of $U(1)_V$ embedded into $SU(N)$ as $\lambda=\text{diag}(\dots,0,1_{i-1},-1,-1,1,0,\dots)$.

Since we are working in regular gauge, we can impose regularity at the origin. More specifically, the fields should be smooth enough to allow their derivative to be defined, which is especially nontrivial given the angle dependence of $\hat{r}$ and $\hat{r}\cdot\tau$. We also assume them to decay to zero at infinity.

For $U(1)_A$ taking the divergence of $A_i$ we get $\left(\frac{2}{r}a_1+a'_1\right)\hat{r}\cdot\tau$; the value at the origin is independent of the direction from which we approach $r=0$ if $a_1=O(r^2)$. $A_0$ is discontinuous unless $a_0(0)=0$, while $a'_0(0)$ remains free: indeed $\partial_iA_0 = \frac{a_0}{r}\tau_i + r\left(\frac{a_0}{r}\right)'\hat{r}_i(\hat{r}\cdot\tau)$, but for $a_0$ linear in $r$ the angle dependent term cancels at $r=0$.

For $U(1)_V$ $(\vec\nabla \tilde A_0)^2=\lambda^2 (\tilde a'_0)^2$, so $\tilde a'_0(0)=0$. On the other hand $\tilde a_0(0)$ remains unconstrained. $\partial_i\tilde{A}_i=\tilde a'_1+\tilde a_1\frac{2}{r}$ imposes $\tilde a_1(0)=0$; conversely $\tilde a'_1(0)$ finite but free guarantees the independence on angles of $\partial_i\tilde A_j=\tilde a'_1\hat{r}_i\hat{r}_j+\frac{\tilde a_1}{r}\left(\delta_{ij}-\hat{r}_i\hat{r}_j\right)$ at $r=0$.

			\section{The Rubakov's calculation in the $\psi\chi\eta$ model}
			\label{sec:rubakov}
		
		We aim to perform a calculation similar to the Rubakov one, with the theory at hand. In 2D we have a two dimensional $U(1)_A\times U(1)_V$ theory, where the gauge fields, $a$ and $v$, are coupled axially and vectorially with the 2D fermions (see the charges in \ref{chargespsieta2d}), 
		\begin{align} \label{eq:the2daction}
			&S^{\rm eff}_{2{\rm D}}=\int drdt \;\; \frac{4\pi}{2g^2_a} r^2 f_{ij}(a)f^{ij}(a) + \frac{4\pi}{2g^2_v} r^2 f_{ij}(v)f^{ij}(v)  \nn \\
		&	\qquad \qquad +\bar{\psi}^m\gamma^i\left(\partial_i+2i\gamma^5 a_i-2iv_i \right)\psi^m + \bar{\eta}^A\gamma^i\left(\partial_i + i\gamma^5 a_i+ iv_i \right)\eta^A
		\end{align}
		where 
		\be
		f_{ij}(a)=\partial_i a_j-\partial_j a_i\;, \quad f_{ij}(v)=\partial_i v_j-\partial_j v_i\;.
		\ee
		accompanied by the boundary conditions 
		\be
		a_0|_{r=0}=0\;, \quad \partial_r a_1|_{r=0}=0\;, \quad \partial_r v_0|_{r=0}=0\;, \quad v_1|_{r=0}=0\;, \ee and 
\be
		\psi^m_l|_{r=0}=\psi^m_r|_{r=0} \;\qquad \eta^A_{l}|_{r=0}=\eta^A_{r}|_{r=0}\;. \label{eq:boundaryconditions}
		\ee

There are a few important observations to make before we start to compute the correlation functions of this theory.
 Being $a$ and $v$ two-dimensional gauge fields, they don't carry any local degrees of freedom. Nevertheless, including them in the computation is important to properly reproduce the Coulombic attraction. The anomaly between $U(1)_A$ and $U(1)_V$ cancels:
\be \sum_{\rm fermions} q_A \cdot q_V = \sum_{I=1, 2} 2 \cdot 2 + \sum_{A=1, ... 8} -1 \cdot 1 = 0 \ee
thus allowing  their simultaneous gauging.  
2D QED in general is not conformal because of the dimensional coupling, but the coupling of this theory, $\frac{4\pi}{g^2}r^2$ does not introduce any new mass scale: scale invariance is preserved.

  The idea is, as in the Rubakov seminal work, to extract the vev of an operator $\mathcal{O}$ from the two-point Euclidean correlation function,  by using cluster decomposition
  \be
 \lim_{t\to \infty} \brc\bar{\mathcal{O}}(r, t)\mathcal{O}(r, 0) \ckt=\brc \bar{\mathcal{O}} \ckt \brc \mathcal{O}\ckt\;.
	   \ee

		\subsection{%
Rewrite the effective action}

		The first step is to rewrite the gauge field, in terms of scalars, 
		\be
		a_i=\epsilon_{ij} \partial_j \rho^a + \partial_i \lambda^a \qquad  v_i=\epsilon_{ij} \partial_j \rho^v + \partial_i \lambda^v\;. 
		\ee
		In the form language, this line is simply
		\be
		a=*d \rho^a + d \lambda^a \quad  v=*d\rho^v + d \lambda^v\;. 
		\ee
		Written in this term, the gauge sector of the Lagrangian reads
		\be
		S^{\rm eff}_{gauge}=\int drdt \;\; \frac{4\pi}{g^2_a} \rho^a (\Box r^2 \Box) \rho^a + \frac{4\pi}{g^2_v}  \rho^v (\Box r^2 \Box) \rho^v
		\ee
		while clearly $\lambda^a$ and $\lambda^v$ drop from the action because of gauge invariance. 
		 
		The boundary conditions on the gauge fields become boundary conditions on the $\rho$ fields:
		\bea
		a_0|_{r=0}=0\;, \quad \partial_r a_1|_{r=0}=0 \quad \implies \quad \partial_r \rho^a|_{r=0}=0\\
		 \partial_r v_0|_{r=0}=0\;, \quad v_1|_{r=0}=0 \quad \implies \quad \partial_0 \rho^v|_{r=0}=0
		\eea
		which are essential to determine the correct propagator to use for the different $\rho$ fields.
		 The upshot is that we should impose Dirichlet boundary conditions on $\rho^v$ and Neumann boundary conditions on $\rho^a$.

		At this point, it is possible to decouple the scalar and the fermion sectors through a proper change of variables. In general one can decouple a fermion $\chi$ coupled with $a$ and $v$
		\be
		\int \bar{\chi}\gamma^i(\partial_i - i q_a\gamma^5 a_i - i q_v v_i)\chi
		\ee
		with the change of variable 
		\be
		\chi=e^{-q_a \rho^a  - q_v \gamma^5 \rho^v}\chi_{\rm free} \qquad \bar{\chi}=\bar{\chi}_{\rm free} e^{q_a \rho^a - q_v \gamma^5 \rho^v}\;.
		\ee

        Notice that the Dirichlet boundary condition for $\rho^v$, $\rho^v|_{r=0}=0$, implies that the vectorial boundary condition holds for the free fields also
      \be
      \chi_l=\chi_r \implies (\chi_{\rm free})_l = (\chi_{\rm free})_r\;. \label{eq:boundaryconditionfree}
      \ee

			In general, this change of variable has a nontrivial Jacobian.
				A way to compute the Jacobians to use the standard axial and vectorial anomalies
	    \be
	    \psi\to e^{i\alpha} \psi \implies \frac{iq_a}{2\pi}\int \alpha da  \qquad \psi\to e^{i\nu \gamma^5} \psi \implies \frac{iq_v}{2\pi}\int  \nu dv  
	    \ee
	    once one performs an axial or vectorial rotation on the field.\footnote{We remind that for our theory the overall $U(1)_A-U(1)_V$ anomaly cancels, because of the charge assignments. However here we are not performing axial and vectorial gauge transformations, but more general ones, so the overall phase doesn't vanish.}
	   		By plugging $\alpha= -i q_a\rho^a$ and $\nu= -i q_v \rho^v$, we obtain
		\bea &&\Delta S= \frac{1}{\pi}\int  q_a^2 \rho^a da + q^2_v \rho^v dv\nn \\ 
	&&=	\frac{1}{2\pi}\int  \left( q^2_a \rho^a (\partial_i a_j-\partial_j a_i) + (a \leftrightarrow v)\right)  dx^idx^j\nn \\ 
	&&=	\frac{1}{\pi}\int  \left( q^2_a \rho^a \partial_i a_j\, \epsilon^{ij}  + (a \leftrightarrow v)\right) d {\rm vol}\nn \\ 
	&&=	-\frac{1}{\pi}\int  \left( q^2_a \rho^a \Box \rho^a + (a \leftrightarrow v)\right)   d  {\rm vol} 
		\eea
		Thus, with multiple fermions, the new contribution for the partition function is
		\be
		\Delta S= -\frac{1}{\pi}\int \mathcal{A} \rho^a \Box \rho^a + \mathcal{V} \rho^v \Box \rho^v
		\ee
		where 
		\be
		\mathcal{A}=\sum q_a^2 \qquad \mathcal{V}=\sum q_v^2 
		\ee
		
		In our theory, eq~(\ref{eq:the2daction}),  we have 
		\be
		\mathcal{A}=\underbrace{2\cdot 2^2}_{\psi}+ \underbrace{8\cdot(1)^2}_{\eta}=16 \qquad \mathcal{V}=\underbrace{2\cdot 2^2}_{\psi}+ \underbrace{8\cdot(-1)^2}_{\eta}=16
		\ee
		so we get the renewed gauge action 
		\be
		S^{\rm eff}_{\rm gauge}=\int drdt \left( \frac{4\pi}{g^2_a} \rho^a (\Box r^2 \Box) \rho^a - \frac{\mathcal{A}}{\pi} \rho^a \Box \rho^a + \left(a \leftrightarrow v\right) \right) 
		\ee

		With the change of variable above we have factorized the computation of any correlation function as the product of a bosonic times a free fermion computation
		\be
		\mathcal{O}_i=e^{{\tilde Q}_{i, A} \rho^a}e^{{\tilde Q}_{i, V} \rho^v}\mathcal{O}_{i, {\rm free}}
		\ee 
		so
		\be
		\brc \mathcal{O}_i \mathcal{O}_j \ckt = \brc e^{{\tilde Q}_{i, A} \rho^a}e^{{\tilde Q}_{j, A} \rho^a} \ckt \brc e^{{\tilde Q}_{i, V} \rho^v}e^{{\tilde Q}_{j, V} \rho^v}\ckt \brc \mathcal{O}_{i, {\rm free}} \mathcal{O}_{j, {\rm free}} \ckt 
		\ee
		Here the $\tilde Q_{A/V}$ are the ones that appear because of the change of variable. Notice that they don't coincide, in general, neither with the axial nor with the vectorial charge of the operators. An operator can be  written  schematically as
		\be
		\mathcal{O}= \psi_\ell^a \psi_r^b \bar{\psi}_\ell^{\bar{a}} \bar{\psi}_r^{\bar{b}} \eta_\ell^c \eta_r^d \bar{\eta}_\ell^{\bar{c}} \bar{\eta}_r^{\bar{d}}
		\ee
		where the upper indices are exponents ( the actual $SU(2)_J$ and $SU(8)$ indices are not shown).  
		We have (the barred fermions give the opposite contribution to $\rho$)
				\beq
    			\centering
    		\begin{tabular}{c|c|c|c|c}
    				& $\psi_\ell$ & $\psi_r$ & $\eta_\ell$ & $\eta_r$ \\
    			\hline
     \ $\tilde Q_A$\phantom{\Big|}  & $-2$ & $-2$ & $-1$ & $-1$ \\
    	 	$\tilde Q_V$ & $-2$ & $2$ & $1$ & $-1$ \\
    	 	$Q_A$ & $2$ & $-2$ & $1$ & $-1$ \\
    		 	$Q_V$ & $2$ & $2$ & $-1$ & $-1$ \\
    		\end{tabular}
		\eeq
				Recollecting we have 
		\bea
	&&	\tilde Q_{\mathcal{O}, A}=-2(a+b)-(c+d) - (\text{barred}) \nn \\
	&&	\tilde Q_{\mathcal{O}, V}=-2(a-b)-(c-d) - (\text{barred})\;.
		\eea

			\subsection{%
		Propagators}

			Here we  want to compute the propagators for the bosonic and  fermion fields  on the half-plane, each  associated with its proper boundary condition. 

		\subsubsection*{$\rho^a$ propagators}

        The action for the field $\rho^a$ is 
       \be
       \int drdt \left( \frac{4\pi}{g^2_a} \rho^a (\Box r^2 \Box) \rho^a - \frac{\mathcal{A}}{\pi} \rho^a \Box \rho^a \right)  \qquad \mathcal{A}=16
       \ee
		accompanied by the Neumann boundary condition at the monopole core
		\be
		\partial_r \rho^a|_{r=0}=0\;.
		\ee
		The propagator is 
		\be
		G=\frac{1}{\left(\frac{4\pi}{e^2} \Box r^2 \Box - \frac{\mathcal{A}}{\pi} \Box  \right)}= -\frac{\pi}{\mathcal{A}} \frac{1}{\left(\Box - \Box \frac{r^2}{k} \Box\right)}
		\ee
		where $k=\frac{\mathcal{A}e^2}{4\pi^2}$.
		by using the formal  property
		\be
		\frac{1}{A}-\frac{1}{B}=\frac{1}{A}(B-A)\frac{1}{B}\;,
		\ee
		with 
		\be
		A=\Box \quad \text{and} \quad B=\Box - \frac{k}{r^2}\;.
		\ee		
		one has
		\be
		G_{\rm eff} = \frac{\pi}{\mathcal{A}} \left[ \Box^{-1} - \left(\Box- \frac{k}{r^2}\right)^{-1} \right] \;.
		\ee
		By using 
		\be
		\Box^{-1}(x, y)= \frac{1}{4\pi} \log(\mu^2|x-y|^2) + {\rm analytic}
		\ee
		and imposing the boundary condition $\partial_r \rho|_{r=0}=0$ one obtains 
		\be
		\Box^{-1}(x, y)= \frac{1}{4\pi} \log(\mu^2|x-y|^2) + \frac{1}{4\pi} \log(\mu^2|x-\bar{y}|^2)\;, 
		\ee
		where $\bar{y}=(t_y, -r_y)$ (the inverse of $y$ respect to the boundary $r=0$).
		
		The green function
		\be
		\left(\Box- \frac{k}{r^2}\right)^{-1}(r_1, t_1, r_2, t_2)=\frac{1}{2\pi} M_{d(k)}\left(1+\frac{(r_1-r_2)^2+(t_1-t_2)^2}{2 r_1 r_2}\right)\;.
		\ee
		where $d(k)=\frac{1}{2}(-1 \pm \sqrt{1+4k})$, and $M_{d(k)}(x)$ is the solution\footnote{The correct function is totally fixed requiring regularity at $z\to \infty$. $M_d(z)$ correspond to $\text{LegendreQ[d, 0, 3,z]}$ in Mathematica, where the argument $3$ specify the correct branch structure we need ($3$ means brunch cut between $-\infty$ and $+1$).} of the Legendre differential equation
		\be
		(1-x^2)M_d''-2xM_d' + d(d+1)M_d=0
		\ee
		which it is regular at $x\rightarrow \infty$, while it is singular at $x=\pm 1$ ($x> 1$ physically, $x\rightarrow 1$ as $r_1, t_1\rightarrow r_2, t_2$). The scaling for $\frac{(t_1-t_2)^2}{r_1 r_2} \gg 1$ is $\Delta t^{-(1+\sqrt{1+4k})}$, therefore this contribution vanishes respect the logarithmic one of the free propagator.

        As we can neglect the contribution from $M_d$ in the $\Delta t\to \infty$ limit, we obtain
        \be
        G_{\rm eff} = \frac{1}{\mathcal{A}}\log(\Delta t)+ ...
        \ee
  
		\subsubsection*{$\rho^v$ propagators}
		
		The action for the field $\rho^v$ is identical to the one for $\rho^a$, but now we have Dirichlet boundary conditions, 
        \be
        \rho^v|_{r=0}=0
        \ee
        As before one can write the propagator as 
        \be
        G_{\rm eff} = \frac{\pi}{\mathcal{V}} \left[ \Box^{-1} - \left(\Box- \frac{k}{r^2}\right)^{-1} \right] \;.
		\ee
        but now we use the Dirichlet condition for the $\Box^{-1}$ part
        \be
		\Box^{-1}(x, y)= \frac{1}{4\pi} \log(\mu^2|x-y|^2) - \frac{1}{4\pi} \log(\mu^2|x-\bar{y}|^2)\;, 
		\ee
        while we are free to use 
        \be
		\left(\Box- \frac{k}{r^2}\right)^{-1}(r_1, t_1, r_2, t_2)=\frac{1}{2\pi} M_{d(k)}\left(1+\frac{(r_1-r_2)^2+(t_1-t_2)^2}{2 r_1 r_2}\right)\;.
		\ee
        as
        \be
        M_{d(k)}\left(1+\frac{(r_1-r_2)^2+(t_1-t_2)^2}{2 r_1 r_2}\right) \sim \left(\frac{(r_2)^2+(t_1-t_2)^2}{2 r_1 r_2}\right)^{-d-1} \propto r_1^{d+1}\to 0\;.
        \ee

        The overall contribution has no logarithmic growth in this case: e.g. by taking $r_1=r_2=R$ and $t_1-t_2=\Delta t$ one gets
        \bea
		&& \Box^{-1}(x, y)= \frac{1}{4\pi} \log(\mu^2\Delta t^2) - \frac{1}{4\pi} \log(\mu^2 (\Delta t+ 2R)^2) \nn \\ 
		&& =\frac{1}{4\pi}\log\left(\frac{\Delta t^2}{(\Delta t+ 2R)^2}\right)=\frac{1}{4\pi}\log\left(1-4\frac{R}{\Delta t}+...\right)=0-\frac{1}{\pi}\frac{R}{\Delta t} + o(R/\Delta t)
        \eea

		\subsubsection*{Free fermion propagators}
		With our conventions, $z=\tau+ i r$  and $\bar{z}=\tau-ir$. By defining 
		\be
		G(z, z_0)=\left(\begin{array}{cc} \langle\bar{\psi}_l(z_0)\psi_l(z)
\rangle & \langle\bar{\psi}_r(z_0)\psi_l(z)
\rangle \\
		\langle\bar{\psi}_l(z_0)\psi_r(z)
\rangle & \langle\bar{\psi}_r(z_0)\psi_r(z)
\rangle \end{array}\right)
		\ee
	    then
		\be
		\left(\begin{array}{cc} 0 & \partial \\ \bar{\partial}  & 0\end{array}\right)\left(\begin{array}{cc} \langle\bar{\psi}_l(z_0)\psi_l(z)
\rangle & \langle\bar{\psi}_r(z_0)\psi_l(z)
\rangle \\
			\langle\bar{\psi}_l(z_0)\psi_r(z)
\rangle & \langle\bar{\psi}_r(z_0)\psi_r(z)
\rangle \end{array}\right)=\left(\begin{array}{cc} 0 & \delta^2(z-z_0)  \\
			\delta^2(z-z_0) & 0 \end{array}\right)
		\ee
		and
		\be
		\left(\begin{array}{cc} \langle\bar{\psi}_l(z_0)\psi_l(z)
\rangle & \langle\bar{\psi}_r(z_0)\psi_l(z)
\rangle \\
			\langle\bar{\psi}_l(z_0)\psi_r(z)
\rangle & \langle\bar{\psi}_r(z_0)\psi_r(z)
\rangle \end{array}\right)\left(\begin{array}{cc} 0 & \partial_0 \\ \bar{\partial}_0  & 0\end{array}\right)^T=\left(\begin{array}{cc} 0 & \delta^2(z-z_0)  \\
			\delta^2(z-z_0) & 0 \end{array}\right)\;.
		\ee
 supplemented by the microscopic boundary conditions \ref{eq:boundaryconditions} (and \ref{eq:boundaryconditionfree}), which we repeat here
       \be 
		\psi_l = \psi_r\;. 
		\ee
  
		{Let's consider the slightly more general boundary condition $\psi_l=e^{i\theta}\psi_r$, keeping in mind that, for our system, we are interested in $\theta \ll 1$. } The Green's function is uniquely determined
  	\be
		\left(\begin{array}{cc} \langle\bar{\psi}_l(z_0)\psi_l(z)
\rangle & \langle\bar{\psi}_r(z_0)\psi_l(z)
\rangle \\
			\langle\bar{\psi}_l(z_0)\psi_r(z)
\rangle & \langle\bar{\psi}_r(z_0)\psi_r(z)
\rangle \end{array}\right)= \frac{1}{2\pi}\left(  \begin{array}{cc} \frac{1}{z-z_0} & \frac{e^{i\theta}}{z-\bar{z}_0} \\ \frac{e^{-i\theta}}{\bar{z}-z_0} & \frac{1}{\bar{z}-\bar{z}_0} \end{array}\right)\;,
		\ee
		as it can be checked explicitly by plugging it into the differential equations.
		
		It is useful for us to expand the Green's function for $t-t_0\gg x, x_0$. For simplicity, let us take $x=x_0=r$. Then
		\be
		G=\frac{1}{2\pi}\left\{\frac{1}{\Delta t} \left(  \begin{array}{cc} 1 & e^{i\theta} \\ e^{-i\theta} & 1 \end{array}\right)+ \frac{2ir}{\Delta t^2} \left(  \begin{array}{cc} 0 & e^{i\theta} \\ e^{-i\theta} & 0 \end{array}\right) +  O\left(\frac{1}{\Delta t^2}
\right)\right\}
		\ee
		
		As the green function determinant expansion for large $\Delta t$
		\be
		\det[G] \propto \frac{4 r^2}{\Delta t^4}+ ...
		\ee
	    the correlations function involving both the two chiralities of the same fermion, 
	    \be
	    \langle(\bar{\psi}_l\bar{\psi}_r...)_z (\psi_l\psi_r...)_z
\rangle \propto \det[G]
	    \ee
		scales as $\frac{1}{\Delta t^4}
$, and not as $\frac{1}{\Delta t^2}$ as naive dimensional analysis might suggest.
				Differently, if two fermions are of different flavors, 
		\be
		\langle(\bar{\psi}_l^1\bar{\psi}_r^2...)_z (\psi_l^1\psi_r^2...)_z
\rangle \propto G_{ll}G_{rr}
		\ee
	  it scales as $\frac{1}{\Delta t^2}$, as expected.

		\subsection{%
			 Operators and   condensate in 2D}

        It is convenient to classify the operators in several classes, 	  schematically as
        \be
        \mathcal{O}= (\psi^I_l \psi^I_r)^k (\psi_{l/r}^I)^{m} (\eta^A_l \eta^A_r)^j (\eta_{l/r}^A)^{n} (\bar{\psi}^I_l \bar{\psi}^I_r)^{\bar{k}} (\bar{\psi}_{l/r}^I)^{\bar{m}} (\bar{\eta}^A_l \bar{\eta}^A_r)^{\bar{j}} (\bar{\eta}_{l/r}^A)^{\bar{n}}\;,
        \ee
         labeled by eight numbers $k, \bar{k}, j, \bar{j}, m, \bar{m} , n, \bar{n} $:
        \begin{itemize}
        	\item The number $0\le k, \bar{k} \le 2$ of couples $\psi^I_l\psi^I_r$ (or $\bar{\psi}^I_l\bar{\psi}^I_r$)  where both the two chiralities of the same $\psi^I$ fermion are presents. Each of them will contribute with a factor $\Delta t^{-4}$ to the correlation function.
        	\item The number $0 \le j, \bar{j} \le 8$ of couples $\eta^A_l\eta^A_r$ (or $\bar{\eta}^A_l\bar{\eta}^A_r$) where both the two chiralities of the same $\eta^A$ fermion are presents. Again, each of them will contribute with a factor $\Delta t^{-4}$ to the correlation function.
        	\item The numbers $0 \le m  \le 2-k$ ($0 \le \bar{m}  \le 2-\bar{k}$) of times a $\psi^I$ ($\bar{\psi}^I$) appears with only one of its chiralities in the operator. 
        	\item The numbers $0 \le n \le 8-j $  ($0 \le \bar{n} \le 8-\bar{j}$) of times a $\eta^A$ ($\bar{\eta}_A$) appears with only one of its chiralities in the operator.
        \end{itemize}
			With these eight numbers, it is possible to determine the scaling of the correlation function for large $\Delta t$, as
			\be
			    \mathcal{O} \propto \Delta t^\beta ,
			\ee 
			 where
		\be
		\beta=\frac{\tilde{Q}^2}{16}- \big(4(k+\bar{k}+ j+\bar{j}) + m+ \bar{m}+ n+ \bar{n}\big)\;,
		\ee
       being 
       \be
       \tilde Q= 4(k-\bar{k})+2(j-\bar{j})+ 2(m-\bar{m})+(n-\bar{n})\;.
       \ee
  	   If $\beta<0$ the operator $\mathcal{O}$ doesn't condense, if $\beta=0$ it condenses, while any value $\beta>0$ doesn't satisfy the cluster property. The task is to search all the value of $k, \bar{k}, j, \bar{j}, m, \bar{m}, n, \bar{n}$ that satisfy $\beta=0$.
  	   
  	   We can see that $\beta$ is maximized if one set either $\bar{k}=\bar{j}=\bar{m}=\bar{n}=0$ or $k=j=m=n=0$. Then only a small number ($\sim 100$) of cases remain, and we can explore all of them by exhaustion. The upshot is that $\beta\le 0$, with $\beta=0$ only for $k=j=m=n=0$ (the identity operator), or $k=2, j=0, m=0, n=8$, corresponding schematically to $\psi^4\eta^8$.
  	   
  	   In general, there are many operators of this form. First of all, it is possible to classify them according to their $U(1)$ axial charge, $Q_A=-8, -6, ...,0,..., 8$, which depend on the number of the left and of the right $\eta^A$ present in the operator. Clearly, at the non-perturbative level, only the uncharged operator can condense as a local operator. Explicitly, the integration on the $\lambda$ mode of the gauge field renders any correlation function of the charged operator vanishing.
  	   
  	   What remains is a vector space of $70$ operators, $V_{4,4}$ with four $\eta^A_l$ and four $\eta^A_r$
  	   \be
  	   (\psi_l^1\psi_r^1)(\psi_l^2\psi_r^2) \eta^{A_1}_l...\eta^{A_4}_l \eta^{B_1}_r...\eta^{B_4}_r\;.
  	   \ee  
  	  
         It is worth noticing that $V_{4,4}$ does not decompose as a direct sum of $SU(8)$ irreps, as it is not closed under the $SU(8)$ action. However, within $V_{4,4}$ there is an operator invariant under $SU(8)$, 
  	   \be
  	   (\psi_l^1\psi_r^1)(\psi_l^2\psi_r^2) \eta^{A_1}_l...\eta^{A_4}_l \eta^{B_1}_r...\eta^{B_4}_r \epsilon_{A_1...B_4}\;.
  	   \ee  

        With what we know till now, we cannot say if $SU(8)$ is broken or not. To answer it is necessary to compute explicitly which subspace of $V_{4,4}$ does not get a vev, by evaluating the combinatorial factor associated with the fermionic Wick contractions.
        
       { 
       From the microscopic analysis of the boundary conditions, (\ref{eq:boundaryconditions}) and (\ref{eq:boundaryconditionfree}), we know that all the $\theta$ parameters vanish in our model, up to corrections due to the finite monopole core size. %
       However, let's keep momentarily a common $\theta_{\eta^1}=...=\theta_{\eta^8}$, which does not affect the result below. }
        
        One can denote an element of a particular basis of $V_{4,4}$ by providing a permutation $\sigma$ %
        \be
        \mathcal{O}_{\sigma}= \psi^4 \eta^{\sigma(1)}_1....\eta^{\sigma(4)}_4 \eta^{\sigma(5)}_2....\eta^{\sigma(8)}_2
        \ee
        Clearly this notation is redundant, as $\mathcal{O}_\sigma= \pm \mathcal{O}_{\sigma'}$ if $(\sigma)^{-1} \sigma'$ acts only  by reshuffling the first $1, ..., 4$ and $5, ..., 8$ among each other, separately.
        
        At this point, we can compute 
        \be\langle\mathcal{O}_{\sigma'}^\dagger \mathcal{O}_\sigma
\rangle= \braket{(\bar{\psi}^4 \bar{\eta}^{\sigma'(8)}_2....\eta^{\sigma'(5)}_2 \eta^{\sigma'(4)}_1....\eta^{\sigma'(1)}_1)_{x_1} (\psi^4 \eta^{\sigma(1)}_1....\eta^{\sigma(4)}_4 \eta^{\sigma(5)}_2....\eta^{\sigma(8)}_2)_{x_2}}\ee

        We can rearrange the fermions by paying a sign
  		\bea
  		(-1)^{|\sigma|}(-1)^{|\sigma'|}\langle\bar{\eta}^{8}_{i_{\sigma'(8)}}...\bar{\eta}^{1}_{i_{\sigma'(1)}}  \eta^{1}_{i_{\sigma(1)}}...\eta^{8}_{i_{\sigma(8)}}
\rangle \nn \\
  		i_1=...=i_4=1 \qquad i_5=...=i_8=2\;.
  		\eea 
  		
  		 Now, let's call $A$ the number of couples $(i_{\sigma'(a)}=1, i_{\sigma(a)}=2)$ and $B$ the couples $(i_{\sigma'(a)}=2, i_{\sigma(a)}=1)$. As there is the same number of left and right fermions both in the $\mathcal{O}$ and in $\mathcal{O'}$, then $A=B$. But the correlation function gives a factor $e^{-(A-B)i\theta}$, so the phase cancels overall.  The upshoot is that 
  		\be\langle\mathcal{O}_{\sigma'}^\dagger \mathcal{O}_\sigma 
\rangle \propto (-1)^{|\sigma|}(-1)^{|\sigma'|}\ee
  
  		This tells us that the unique operator that can condense can be written as
  		\be
  		\sum_\sigma (-1)^{|\sigma|} \mathcal{O}_\sigma
  		\ee
  		which is the $SU(8)$ invariant one, 
  		\be
  		\psi^4 \eta_1^{A_1}...\eta_1^{A_4}\eta_1^{B_1}...\eta_1^{B_4}\epsilon_{A_1...A_4B_1...B_4}\;.
  		\label{condensate11}
  		\ee

  		Summarizing, by computing the dynamics of the IR effective theory we confirm that all the global symmetries of the theory remain unbroken by the condensate. Notice that this is a non-trivial result, as we might have found a non-vanishing vev for operators that breaks, e.g. $SU(8)$ or rotations. It is worth noticing that the operator that condenses has the same structure as the 't Hooft operator, which captures the effect of the anomaly in IR. 
  
		\subsection{%
Uplift the condensate in 4D}
		
		We have pinned down exactly the operator that condenses in the  2D  EFT.
		 We have now to uplift the result to 4D, and determine which operators condense. To do so, one can simply remember that for a positively charged fermion
		\be
		\psi_{4{\rm D}}=\frac{1}{r} \psi_l^m \Omega_{\mu, \mu-1/2, m}  + \left(j>\mu-1/2\right)\;,
		\ee
		while for a negatively charged particle
	    \be
	    \psi_{4{\rm D}}=\frac{1}{r} \psi_r \left(\sigma^2 \Omega_{-\mu, -\mu-1/2, m}\right)+ \left(j>-\mu-1/2\right)\;.
	    \ee
	
		In our case 
		\bea
		&& (\psi^{i, i})^2 (\psi^{i, i})^2 \eta^{A_1}_{i+1}...\eta^{A_4}_{i+1} \eta^{B_1}_{i}...\eta^{B_4}_{i} \epsilon_{A_1...A_4B_1...B_4} = \nn \\
		 &&\qquad  \qquad = \frac{1}{r^{12}} \psi_l^{1/2} \Omega_{1, 1/2, 1/2} \psi_l^{-1/2} \Omega_{1, 1/2, -1/2}    \psi_r^{1/2} \left(\sigma^2 \Omega_{1, 1/2, 1/2}\right) \psi_r^{-1/2} \left(\sigma^2 \Omega_{1, 1/2, -1/2}\right)\cdot \nn \\
		&& \qquad  \qquad \quad  \cdot \, \eta^{A_1}_{1} \Omega_{1/2, 0, 0} ...\,\eta^{A_4}_{1} \Omega_{1/2, 0, 0} \eta^{B_1}_{1} \left(\sigma^2\Omega_{1/2, 0, 0}\right) ...\, \eta^{B_4}_{1} \left(\sigma^2\Omega_{1/2, 0, 0}\right)\;.
		\eea
		
		Here, for simplicity, we have suppressed the Lorentz index of every monopole spinor spherical harmonics. Clearly, because of all these Lorentz indices, it is possible to form many tensors, all of which might condense. 
		
		To understand the Lorentz structure, notice:
		\begin{enumerate}
			\item 	Because of statistics, the $\psi$s ought to be contracted to form a Lorentz singlet,  
			\be\psi^{i}_\alpha\psi^{i}_\beta \epsilon^{\alpha\beta}\psi^{i}_\gamma\psi^{i+1}_\delta \epsilon^{\gamma\delta}\ee
			\item   The condensate is by construction symmetric under exchange of any of the spinorial indices of the $\eta^i$s and of the $\eta^{i+1}$s.  This means 
			\be
			\eta^{A_1}_{\{\alpha_1}....\eta^{A_4}_{\alpha_4\}}
			\ee
			and forms two $(2, 0)$ Lorentz representations.
			\item By combining them one obtains
			\be(2, 0)\otimes (2, 0)= (0, 0) \oplus (1, 0) \oplus ... \oplus (4, 0).\ee
			 All of these operators condense.
		\end{enumerate}

		It is not difficult to give an explicit construction to the first two of these operators and compute the condensates
		\bea%
	&&	\langle\mathcal{O}_{(0, 0)}\rangle=\langle(\psi^i \epsilon \psi^i) (\psi^{i+1} \epsilon \psi^{i+1}) (\eta^{A_1}_i \epsilon \eta^{B_2}_{i+1})...(\eta^{A_7}_i \epsilon \eta^{B_8}_{i+1})\epsilon_{A_1...B_8}
\rangle\propto\frac{1}{\color{red} r^{18}} \;\;, \nn \\%
	&&	\langle\mathcal{O}_{(1, 0)}
\rangle=\langle(\psi^i \epsilon \psi^i) (\psi^{i+1} \epsilon \psi^{i+1}) (\eta^{A_1}_i \bar{\sigma}^{[\mu}\sigma^{\nu]} \eta^{A_2}_{i+1})...(\eta^{A_7}_i \epsilon \eta^{A_8}_{i+1})\epsilon_{A_1...B_8}
\rangle\propto\frac{1}{\color{red} r^{18}} \left(\begin{array}{cc}
			0 & i \hat{x} \\
			-i \hat{x} & \epsilon_{ijk}\hat{x}^k
		\end{array}\right)\;,%
		 \nn \\
	    \eea
		In principle, one can compute the rest of the operators, but they become more cumbersome to write down explicitly.

		It is worth stressing that, even if the condensates have a nontrivial Lorentz tensor structure, overall it does not break the rotational symmetry, as it is clear from the explicit expression of the $(1, 0)$ vev.

\section{Bosonization and scattering  in the $\psi\chi\eta$ model} 
\label{sec:bosonization}

The theory of a Dirac massless 2D fermion,  $\Psi=\left(\begin{array}{c}
	\chi_l \\
	\chi_r
\end{array}\right)$, $ \bar{\Psi}=\left( 
\bar{\chi}_r \  \bar{\chi}_l
\right)$
\be
S_D =\int dx^2 \; \bar{\Psi} \gamma^i \partial_i \Psi=\int dx^2 \left\{\chi_l^\dagger \partial_- \chi_l + \chi_r^\dagger \partial_+ \chi_r\right\} \ ,  \qquad \partial_\pm=\partial_t \pm \partial_x \ ,
\ee
is equivalent to the theory of a single (massless) real boson periodic with $\Phi \sim \Phi+2\pi$ 
\be
S_B=\frac{1}{8\pi}\int dx^2 \partial_i \Phi \partial^i \Phi
\ee

In particular, as $\partial_-\bar{\psi}_l\psi_l=\partial_+\bar{\psi}_r\psi_r$, one can define the $U(1)_l$ and the $U(1)_r$ currents in the fermionic system
\bea
&& U(1)_l: \quad J_l^\mu= \left(\begin{array}{c} \chi_l^\dagger \chi_l\\ -\chi_l^\dagger \chi_l \end{array}\right)\;, \quad  \partial_\mu J_l^\mu =0\;,\nonumber \\ 
&& U(1)_r: \quad  J_r^\mu= \left(\begin{array}{c} \chi^\dagger_r \chi_r\\  \chi_r^\dagger \chi_r \end{array}\right)\;, \quad  \partial_\mu J_r^\mu=0\;.
\eea

Then, thanks to the equation of motions for $\Phi$, $\partial_+\partial_- \Phi=\partial_-\partial_+\Phi=0$, we can define a mapping
\be
\chi^\dagger_l \chi_l = \partial_- \Phi \quad \chi^\dagger_r \chi_r= \partial_+ \Phi
\ee

The dynamics of the scattering of the s-wave sector of a single Dirac fermion can be understood in the bosonized theory as the boundary condition for the field $\Phi$.

 After bosonization we ignore the gauge degrees of freedom and treat the system as a bunch of free bosons.  
We assume that  all symmetries apart from the anomalous one should be preserved. We thus have a set of symmetry-preserving boundary conditions. 
The crucial aspect here is that the chiral theory under consideration is locally vectorial. So once we neglect the gauge fields we can just perform the bosonization in the usual way, as if it were a vectorial model. 
The  2D Dirac fermions (\ref{dirac2dpsieta}) can be traded with a 2D compact bosons $\phi_{\psi}$ and $\phi_{\eta}$ with the same charges of the corresponding fermions with respect to $U(1)_A$, $U(1)_V$, $SU(2)_J$ and $SU(8)_{\eta}$ given in (\ref{chargespsieta2d}).

We search for symmetry preserving boundary conditions using the bosonized picture. The conditions that all nonanomalous symmetries are preserved is thus 
\bea 
&	U(1)_A &\implies\quad  2  \sum_I \partial_0 \phi^I_\psi + \sum_A \partial_0 \phi^A_\eta=0 \nn \\
&	U(1)_V &\implies \quad 2  \sum_I \partial_1 \phi^I_\psi - \sum_A \partial_1 \phi^A_\eta=0 \nn \\
&	SU(2)_J &\implies\quad  \partial_1 \phi_\psi^1-\partial_1 \phi_\psi^2=0 \phantom{\sum_A} \nn \\
&	SU(8)_\eta &\implies \quad  \partial_1 \phi^1_\eta=\partial_1 \phi^2_\eta=\dots=\partial_1 \phi^8_\eta 
\eea
Clearly, these $10$ equations fix totally the dynamics. To solve it in general it is convenient to consider a generic solution

\be
\phi^I_\psi=g^I_\psi(t+r)+h^I_\psi(t-r) \quad \phi^A_\eta=g^A_\eta(t+r)+h^A_\eta(t-r) \label{eq:solutions}
\ee
We have to specify the initial conditions of the scattering problem. To describe the scattering of some incoming particles that arrives at the monopole at time $t_0$, we should impose on the functions $g$ and $h$ that
\be
g^{\#}_{\#}(t)=h^{\#}_{\#}(t)=0 \quad \text{for}\;\;t<t_0\;. \label{eq:scatteringcond}
\ee

By plugging the solutions (\ref{eq:solutions}) into the boundary conditions, and integrating using (\ref{eq:scatteringcond})  we obtain
\bea 
&	U(1)_A &\implies \quad 2  \sum_I g^I_\psi + \sum_A g^A_\eta=- 2  \sum_I h^I_\psi - \sum_A h^A_\eta \nn \\
&	U(1)_V &\implies \quad 2  \sum_I g^I_\psi - \sum_A g^A_\eta= 2  \sum_I h^I_\psi - \sum_A h^A_\eta \nn \\
&	SU(2)_J &\implies \quad  g^1_\psi-g^2_\psi=h^1_\psi-h^2_\psi \phantom{\sum_A} \nn \\
&	SU(8)_\eta &\implies \quad  g^1_\eta-h^1_\eta=g^2_\eta-h^2_\eta=\dots=g^8_\eta-h^8_\eta
\eea
or, in matrix form
\bea
\small
\underbrace{\left(
	\begin{array}{cc|cccccccc}
		2 & 2 & 1 & 1 & 1 & 1 & 1 & 1 & 1 & 1 \\
		2 & 2 & -1 & -1 & -1 & -1 & -1 & -1 & -1 & -1 \\
		\hline 
		1 & -1 & 0 & 0 & 0 & 0 & 0 & 0 & 0 & 0 \\
		0 & 0 & 1 & -1 & 0 & 0 & 0 & 0 & 0 & 0 \\
		0 & 0 & 0 & 1 & -1 & 0 & 0 & 0 & 0 & 0 \\
		0 & 0 & 0 & 0 & 1 & -1 & 0 & 0 & 0 & 0 \\
		0 & 0 & 0 & 0 & 0 & 1 & -1 & 0 & 0 & 0 \\
		0 & 0 & 0 & 0 & 0 & 0 & 1 & -1 & 0 & 0 \\
		0 & 0 & 0 & 0 & 0 & 0 & 0 & 1 & -1 & 0 \\
		0 & 0 & 0 & 0 & 0 & 0 & 0 & 0 & 1 & -1 \\
	\end{array}\right)}_{{\cal A}} \, \left(\begin{array}{c} g^1_\psi \\ g^2_\psi \\ \hline  g^1_\eta \\ g^2_\eta \\ g^3_\eta \\ g^4_\eta \\ g^5_\eta \\ g^6_\eta \\ g^7_\eta \\ g^8_\eta \end{array}\right)= \qquad \qquad  \qquad \qquad  \nn\\ 
\small \qquad \qquad  \qquad  \qquad = \underbrace{\left(
	\begin{array}{cc|cccccccc}
		-2 & -2 & -1 & -1 & -1 & -1 & -1 & -1 & -1 & -1 \\
		2 & 2 & -1 & -1 & -1 & -1 & -1 & -1 & -1 & -1 \\
		\hline
		1 & -1 & 0 & 0 & 0 & 0 & 0 & 0 & 0 & 0 \\
		0 & 0 & 1 & -1 & 0 & 0 & 0 & 0 & 0 & 0 \\
		0 & 0 & 0 & 1 & -1 & 0 & 0 & 0 & 0 & 0 \\
		0 & 0 & 0 & 0 & 1 & -1 & 0 & 0 & 0 & 0 \\
		0 & 0 & 0 & 0 & 0 & 1 & -1 & 0 & 0 & 0 \\
		0 & 0 & 0 & 0 & 0 & 0 & 1 & -1 & 0 & 0 \\
		0 & 0 & 0 & 0 & 0 & 0 & 0 & 1 & -1 & 0 \\
		0 & 0 & 0 & 0 & 0 & 0 & 0 & 0 & 1 & -1 \\
	\end{array}
	\right)}_{{\cal B}} \, \left(\begin{array}{c} h^1_\psi \\ h^2_\psi \\ \hline h^1_\eta \\ h^2_\eta \\ h^3_\eta \\ h^4_\eta \\ h^5_\eta \\ h^6_\eta \\ h^7_\eta \\ h^8_\eta \end{array}\right)
\eea
which allows to write 
\beq
{\bf h}={\cal B}^{-1}  {\cal A} \,\, {\bf g}= {\cal S}  \,\, {\bf g}\;,
\eeq
where $\mathcal{S}$ is the scattering matrix given by
\be 
\mathcal{S} = {\cal B}^{-1}  {\cal A}  = \left(
\begin{array}{cc|cccccccc}
	\frac{1}{2} & -\frac{1}{2} & -\frac{1}{4} & -\frac{1}{4} & -\frac{1}{4} & -\frac{1}{4} & -\frac{1}{4} & -\frac{1}{4} & -\frac{1}{4} & -\frac{1}{4} \\
	-\frac{1}{2} & \frac{1}{2} & -\frac{1}{4} & -\frac{1}{4} & -\frac{1}{4} & -\frac{1}{4} & -\frac{1}{4} & -\frac{1}{4} & -\frac{1}{4} & -\frac{1}{4}  \\  
	\hline
	-\frac{1}{4} & -\frac{1}{4} & \frac{7}{8} & -\frac{1}{8} & -\frac{1}{8} & -\frac{1}{8} & -\frac{1}{8} & -\frac{1}{8} & -\frac{1}{8} & -\frac{1}{8} \\
	-\frac{1}{4} & -\frac{1}{4} & -\frac{1}{8} & \frac{7}{8} & -\frac{1}{8} & -\frac{1}{8} & -\frac{1}{8} & -\frac{1}{8} & -\frac{1}{8} & -\frac{1}{8} \\
	-\frac{1}{4} & -\frac{1}{4} & -\frac{1}{8} & -\frac{1}{8} & \frac{7}{8} & -\frac{1}{8} & -\frac{1}{8} & -\frac{1}{8} & -\frac{1}{8} & -\frac{1}{8} \\
	-\frac{1}{4} & -\frac{1}{4} & -\frac{1}{8} & -\frac{1}{8} & -\frac{1}{8} & \frac{7}{8} & -\frac{1}{8} & -\frac{1}{8} & -\frac{1}{8} & -\frac{1}{8} \\
	-\frac{1}{4} & -\frac{1}{4} & -\frac{1}{8} & -\frac{1}{8} & -\frac{1}{8} & -\frac{1}{8} & \frac{7}{8} & -\frac{1}{8} & -\frac{1}{8} & -\frac{1}{8} \\
	-\frac{1}{4} & -\frac{1}{4} & -\frac{1}{8} & -\frac{1}{8} & -\frac{1}{8} & -\frac{1}{8} & -\frac{1}{8} & \frac{7}{8} & -\frac{1}{8} & -\frac{1}{8} \\
	-\frac{1}{4} & -\frac{1}{4} & -\frac{1}{8} & -\frac{1}{8} & -\frac{1}{8} & -\frac{1}{8} & -\frac{1}{8} & -\frac{1}{8} & \frac{7}{8} & -\frac{1}{8} \\
	-\frac{1}{4} & -\frac{1}{4} & -\frac{1}{8} & -\frac{1}{8} & -\frac{1}{8} & -\frac{1}{8} & -\frac{1}{8} & -\frac{1}{8} & -\frac{1}{8} & \frac{7}{8} \\
\end{array}
\right)
\label{Smatrix}
\ee
From the unitary matrix $\mathcal{S}$ one can read the scattering. Here we give some example 
\begin{itemize}
	\item Sending in a single $\eta^1_{i+1}$ particle is represented as having a kink in $g$
	\be g(t<t_0)=0 \quad g(t\gg t_0) \to 1 \ee
	Using the $\mathcal{S}$ matrix we known that for $h$ the kink is
	\be
	{\bf h}^t (t\gg t_0)=\left( -\frac{1}{4},-\frac{1}{4} \ \vline \  \frac{7}{8},-\frac{1}{8},-\frac{1}{8},-\frac{1}{8},-\frac{1}{8},-\frac{1}{8},-\frac{1}{8},-\frac{1}{8} \right)
	\label{heta}
	\ee
	which is a state with a fractional occupation number.
	\item Scattering a single $\psi_{i}$ particle (with $m=1/2$) we obtain
	\be
	{\bf h}^t (t\gg t_0)=\left( \frac{1}{2},-\frac{1}{2}\  \vline    -\frac{1}{4},-\frac{1}{4},-\frac{1}{4},-\frac{1}{4},-\frac{1}{4},-\frac{1}{4},-\frac{1}{4},-\frac{1}{4} \right)
	\label{hpsi}
	\ee
\end{itemize}
This type of fractional kinks, or microtons,  as outcome of the scattering process is like the one first observed in  \cite{Callan:1983tm} in the context of the GUT standard model monopole. 

In the eigendecomposition of the matrix $\mathcal{S}$ there is a single $-1$ eigenvalue, relative to the state 
\be
{\bf g}_-^t= \left(2, 2, 1, 1, 1, 1, 1, 1, 1, 1\right)\;, \qquad \mathcal{S}\,\, {\bf g}_-=- {\bf g}_-
\ee
while $\mathcal{S}=\mathbbm 1$ on the orthogonal subspace, which we call $\mathcal{H}_0$. Notice that the $\mathcal{H}_0$ spans the subspace of states that are uncharged under $U(1)_A=U(1)_i$.    

The presence of a $-1$ eigenvalue is related to the condensate~(\ref{condensate11}). Indeed there is a multi-particle scattering process, entirely in the Fock space, that is directly related to the condensate. If we take as state the IN-going part of the condensate, we have
\beq
{\bf h}^t = (1,1 |1,1,1,1,0,0,0,0)
\eeq
applying the scattering matrix (\ref{Smatrix}) we obtain
\beq
{\bf g}^t = (-1,-1 | 0,0,0,0,-1,-1,-1,-1)
\eeq
These are exactly the conjugate of the OUT states in the condensate. This is precisely what happens for the simple case of the CR effect in. 
Another way of writing this process is 
\beq
 \psi_{i}^1  + \psi_{i}^2 + \eta_{i+1}^1   + \eta_{i+1}^2   + \eta_{i+1}^3   + \eta_{i+1}^4  \longrightarrow \bar{\psi}_{i+1}^1 +  \bar{\psi}_{i+1}^2 + \bar{\eta_{i}}^5 + \bar{\eta}_{i}^6 + \bar{\eta}_{i}^7 +\bar{\eta}_{i}^8
\label{multiscatanomaly} 
\eeq

This concludes the list of elementary scattering processes which can be readily interpreted in terms of particles: there are the ones where the IN state is neutral under $U(1)_A$, and the one related to the condensate. The most general process we can understand can be decomposed into a superposition (with integer coefficients) of these processes. All other processes involve fractional charges, as we showed in the examples~(\ref{heta},\ref{hpsi}).
\medskip

The problem of fractional OUT states has been known since the first paper by Callan about GUT monopoles \cite{Callan:1983tm}. It also appears for vectorial models with multiple flavors. The GUT problem can be reduced to the $SU(2)$ monopole with $4$ Weyl fermion doublets. Here the fractional number that appears is $\frac{1}{2}$, and these fractional hypothetical states are also called semitons.   This problem is still open and currently debated in the literature. 
The present work presented a new chiral model, the $\psi\chi\eta$, where the same issue of fractionalization appears in the final scattering matrix. We thus arrived at the same old problem in a new model. 
Here we give a brief overview of the possible solutions, with no ambition for the moment to find a definite answer.
Each proposal has some weak points, or at least some not completely proven assumptions. 
\smallskip

In the original paper \cite{Callan:1983tm} it was proposed a sort of density matrix type interpretation. The fractional number could be a probabilistic coefficient. This would require some explanation of how a pure IN state comes as a mixed state. It may be entangled with other degrees of freedom. A measure would project in a certain state in the Fock space of OUT particles and this would then break some of the global symmetries, unless the entanglement also enters into play. 
In \cite{Polchinski:1984uw} it was suggested that the OUT state may be considered as a vacuum pulse, a sort of excitation around the vacuum that cannot be interpreted as a multiparticle state in a Fock space. 
\smallskip

A recent idea has been presented in \cite{Csaki:2021ozp}, following earlier work \cite{Csaki:2020yei,Csaki:2020inw}, see also on a similar line \cite{Khoze:2023kiu}. These works question the fact that at low energy we can neglect higher spin excitations. Using the pairwise helicity formalism, they constructed a state made of OUT multifermions that preserves all the charges. %
To construct this state it is important to include higher angular momentum states, all entangled with the field angular momentum to have a total $J=0$ state.  
Usually, higher angular momentum states are not expected to be relevant in a low-energy scattering setup because they are kept away from the monopole core by the centrifugal barrier (see our discussion in sec.\ref{sec:dirac}). 
The interaction between opposite charges particles would lower the energy but, at small coupling, we do not expect to completely cancel the centrifugal barrier.  So these processes are expected to remain ''below the barrier'' from radius $R_{\rm UV}$ up to radius $1/\epsilon$, where $\epsilon$ is the energy of the incoming particle, thus we expect them to be suppressed at low energy.  Another problem is that this approach does not explain the scattering if we consider the just $1+1$ theory. There must be a solution for the scattering problem even if we take just the $1+1$ theory in isolation, so without higher spin states by definition.  
\smallskip

In \cite{Brennan:2021ewu} it was suggested that the missing charge may be go in the form of soft fermionic radiation. 
The authors of \cite{Hamada:2022eiv}, 
by using the rotor model, suggest the possibility of  excite the dyonic degrees of freedom, even at low energy. Another approach that involved the change in the monopole-dyon is \cite{Brennan:2023tae}. Here it is suggested that a dyon can be created at parametrically small energies. This would be the case if the fermion has JR zero modes and may be an interesting possibility to explore for those cases we excluded in our derivation.  
In \cite{Kitano:2021pwt} they suggested the existence of new fermionic states mas as solitonic excitations, ''pancakes'' of the condensates.  This may reflect the fact that going to higher spin there are multiple possible states with the correct quantum numbers, but it is not clear how to compute the scattering probability.
\smallskip

Another recent idea has been presented in \cite{vanBeest:2023dbu}. Here it is suggested that a topological line form after the scattering and connects the monopole with the OUT states.  This is somehow in line with the original idea of the vacuum pulse of \cite{Polchinski:1984uw}. Bosonization gives a definite answer for the OUT scattering state in $1+1$. It is a state that cannot be interpreted as a single or multi-particle in the Fock space and may be a nonlocal excitation. This type of approach is meant to work also in the isolated $1+1$ theory. It remains difficult to interpret the result in terms of what a physical detector would measure out of a scattering experiment.

To be more precise, the authors of \cite{vanBeest:2023dbu} suggest that we should take the answer from bosonization seriously and, considering the final state of the fermionic theory as an exponential of bosonic operators, they split it into a topological operator connecting the monopole with elementary outgoing fermions (see for example their equations (2.17,2.18) ). The pleasing feature of this approach is that it interprets the OUT state as elementary outgoing fermionic operators, but acting on a \textquotedblleft twisted vacuum\textquotedblright{} that carries non-zero charge. The split into topological line and elementary fermionic excitation is uniquely defined by the condition that the mass dimension of the endpoint of the topological operator should be the lowest possible, or equivalently that in radial quantization the energy is minimized, and the result of the twist operator on the vacuum can be interpreted as the vacuum of a twisted sector of the Hilbert space.

Applying their proposal to the $\psi\chi\eta$ model, we get very simple results. If we send in a $\psi$ particle, the outcome (\ref{hpsi}) is to be interpreted as a pure topological line (i.e. the mass dimension is already minimized). On the contrary, if we send a $\eta$ particle in, the resulting (\ref{heta}) state is to be interpreted as an outgoing anti-$\eta$ of the same flavor, attached to a twist operator. The twist operator is described in the bosonized picture by the fractional numbers
\begin{equation}
    \left( -\frac{1}{4},-\frac{1}{4} \ \vline \  -\frac{1}{8},-\frac{1}{8},-\frac{1}{8},-\frac{1}{8},-\frac{1}{8},-\frac{1}{8},-\frac{1}{8},-\frac{1}{8} \right)\;.
\end{equation}

The line operator acts on the right-moving sector, 
\begin{equation}
    \psi^{i+1}\to e^{\frac{2\pi i}{4}}\psi^{i+1}\;, \quad \eta_i^A \to e^{\frac{2\pi i}{8}} \eta_i^A \;, \label{eq:transformation}
\end{equation}
while it leaves the left-movers, $\psi^{i}$ and $\eta^A_{i+1}$, untouched. Unsurprisingly,  \ref{eq:transformation} is a combination of an $SU(8)$ center, a $\mathbbm Z_{16} \subset U(1)_i$ and a $\mathbbm Z_{16} \subset \tilde U(1)$ transformations\footnote{At is is, this transformation seems to generate a $\mathbbm Z_{16}$ group. However, by applying $8$ times this transformations one obtains the diagonal element of $\mathbbm Z_2 \times \tilde{\mathbbm  Z_2}$, where $\mathbbm Z_2\subset U(1)_i$ and $\tilde{\mathbbm  Z_2}\subset \tilde U(1)$, which acts trivially. In other words only $\frac{U(1)_i\times \tilde U(1)}{\mathbbm Z_2}$ acts faithfully.}
\be
e^{\frac{2\pi i}{8}} \in SU(8)\;, \quad
e^{-\frac{2\pi i}{16}} \in U(1)_i\;, \quad 
e^{\frac{2\pi i}{16}} \in \tilde U(1)\;.
\ee

This is natural, as the topological line can end topologically on the monopole only if it fits on the group preserved by the monopole boundary condition.

\medskip

There is still no definite answer for this unitarity problem, which also appears in the $\psi\chi\eta$ model. In general, it is quite difficult to have an explanation that works also in the reduces $1+1$ model and at the same time can give an answer to what a physical device would measure after the scattering. %

\section{Conclusion  \label{sec:conclude} }

In this work, we approached the Callan-Rubakov problem in chiral theories by analyzing in detail an explicit example. In particular, noticing that for the cases of vector-like theories the UV completion has some role in the explicit computation by Callan and Rubakov, we identified a subset of all the possible Abelian gauge theories that can possess a simple interpretation in terms of the abelianized phase of a UV complete model: the \textit{locally vector-like} models. 

Within this class, we found a simple example which is the IR description of an $SU(N)$ chiral gauge theory: the $\psi\chi\eta$ model.

Thanks to the locally vector-like nature of the model, and the explicit UV completion of the Dirac monopole in terms of a 't Hooft-Polyakov monopole, many of the techniques developed by Callan and Rubakov can be used with little tweaks. In particular:
\begin{itemize}

     \item The explicit UV completion, and the fact that the model abelianizes at a perturbative scale, allows us to understand the IR theory and the magnetic monopole spectrum. In particular, we can ensure that there are no Jakiew-Rebbi zeromodes on the background of the monopoles considered here, thus no degenerate dyons. 

    \item Then, following the example of Rubakov, we computed the Callan-Rubakov condensate around the monopole in the limit of small monopole and small electric coupling. This computation allows us to understand how the symmetry group is broken. We found that only a single operator condenses, roughly $\psi^4 \eta^8$, breaking only the anomalous symmetries and leaving the global symmetry group inherited from the UV completion intact.

    \item Lastly, with this knowledge, we rewrote the effective theory that describes the s-wave fermions in terms of scalars, and here we described the scattering in terms of an S-matrix, completely determined by the unbroken symmetries alone. Our final states are composed of purely outgoing s-wave modes.
\end{itemize}

As for the case of QED with $N_f\ge 4$ or higher-charge fermions, our understanding is not yet satisfactory. The scattering of a single IN fermion leads to an OUT state with a fractional number of fermions, called semitons (or better, microtons), whose interpretation is still very debated.

An analysis similar to ours can be applied to the ''locally vectorial'' theories. However, many other chiral theories such as the one considered in \cite{Smith:2019jnh,Smith:2020nuf}, and whose prototype is (\ref{chiral1a}), do not fall into this category. A natural further step in studying the chiral theories would be to try to tackle these theories by using other classes of UV completions.

\appendix

\section{Conventions}
		
		In the following we will be using the mostly plus convention for the metric, $\eta^{\mu\nu}={\rm diag}(-1, 1, 1, 1)$, and the definition ${\gamma^\mu, \gamma^\nu}=-2\eta^{\mu\nu}$. In our conventions, the Minkowisky 2D gamma matrices (in the chiral basis) are  
		\be
		\label{eq:gamma_matrices_2d_lorentz}
		\gamma^0=\left(\begin{matrix}
			0 & 1\\
			1 & 0
		\end{matrix}\right) \quad \gamma^1=\left(\begin{matrix}
			0 & 1\\
			-1 & 0
		\end{matrix}\right) \quad \gamma^5=\gamma^0\gamma^1=\left(\begin{matrix}
			-1 & 0\\
			0 & 1
		\end{matrix}\right)\;. 
		\ee
		so a left-handed 2D fermion $\gamma^5 \psi_l=-\psi_l$, is left-moving $(\partial_0-\partial_1)\psi_l=0$
		
		In euclidean signature, we use the representation
		\be
		\label{eq:gamma_matrices_2d}
		\gamma^0=\left(\begin{matrix}
			0 & 1\\
			1 & 0
		\end{matrix}\right) \quad \gamma^1=\left(\begin{matrix}
			0 & -i\\
			i & 0
		\end{matrix}\right) \quad \gamma^5=\gamma^0\gamma^1=\left(\begin{matrix}
			-1 & 0\\
			0 & 1
		\end{matrix}\right)\;.
		\ee
		so the massless 2D Dirac equation reads 
		\be
		\left(\begin{matrix}
			0 & \partial\\
			\bar{\partial} & 0
		\end{matrix}\right)\left(\begin{array}{c}
		\psi_l \\ \psi_r
	\end{array}\right)=0\;.
		\ee
		By continuing a purely left-handed (left-moving) fermion, $\psi_l(t+r)$, to the euclidean slice, $t\longrightarrow -i \tau$, one obtains an holomorphic fermion $\psi(-i(\tau + ir))$, which is fully consistent with the 2D equation.	
		
		It is useful to report here a couple of identities that are used multiple times. In euclidean signature the vector and the axial currents 
		\be
		J^i_V=\bar{\psi}\gamma^i\psi =\left(\begin{array}{c}
			\bar{\psi}_l \psi_l + \bar{\psi}_r\psi_r \\
			i(\bar{\psi}_l \psi_l - \bar{\psi}_r\psi_r) \\
		\end{array}  \right)  \quad 
	J^i_A=\bar{\psi}\gamma^i\gamma^5\psi =\left(\begin{array}{c}
		-\bar{\psi}_l \psi_l + \bar{\psi}_r\psi_r \\
		-i(\bar{\psi}_l \psi_l + \bar{\psi}_r\psi_r) \\
	\end{array}  \right) 
		\ee
		are related 
		\be
		J^i_V=i \epsilon^{ij}J^j_A \quad J^i_A=i \epsilon^{ij}J^j_V\;. \label{eq:vectoraxialidentity}
		\ee

		Similarly, we define the Minkowisky $4$D gamma matrices in the chiral basis as
		\be
		\label{eq:gamma_matrices_4d}
		\gamma^0=\left(\begin{matrix}
			0 & 1\\
			1 & 0
		\end{matrix}\right) \quad \gamma^i=\left(\begin{matrix}
			0 & \sigma^i \\
			\bar{\sigma}^i & 0
		\end{matrix}\right) \quad \gamma^5=\left(\begin{matrix}
			-1 & 0\\
			0 & 1
		\end{matrix}\right)\;.
		\ee

\section{Monopole Spherical Harmonics}
\label{appendix:monopole_harmonics}
To exploit spherical symmetry, we would like to decompose the fermion into angular momentum eigenmodes. However the rotation generator is $\tilde L$ and not $L$, so it is convenient to first define the scalar monopole spherical harmonics, $Y_{\mu, j, m}(\theta, \phi)$, with 
		\be j =\mu, \mu+1, ... \quad \text{and} \quad m=-j, -j+1, ... , j-1 , j\ee
		such that 
		\be
		{\tilde L}^2 Y_{\mu, j, m}(\theta, \phi)=j(j+1)Y_{\mu, j, m}(\theta, \phi)\;, \quad \tilde L_z Y_{\mu, j, m}(\theta, \phi)=mY_{\mu, j, m}(\theta, \phi)
		\ee
		as the ordinary spherical harmonics.
An explicit expression for $Y_{\mu, j, m}$ is
		\be
		Y_{\mu, j, m}(\theta, \phi) = N_{\mu, j, m} e^{i \phi  (\mu +m)} (1-\cos(\theta))^{-\frac{\mu +m}{2}} (1+\cos(\theta))^{-\frac{\mu-m}{2}} P_{j+\mu }^{(-m-\mu ,m-\mu )}(\cos(\theta))
		\ee
		where $N_{\mu, j, m}$ is a normalization constant.

We introduce the   spinor monopole spherical  harmonics, 
	\be 
	\Phi^{(1)}_{\mu, j, m}(\theta,\phi)=\left(\begin{array}{c}
		\sqrt{\frac{j+m}{2j}}Y_{\mu,j-\frac{1}{2},m-\frac{1}{2}}(\theta,\phi)\\
		\sqrt{\frac{j-m}{2j}}Y_{\mu,j-\frac{1}{2},m+\frac{1}{2}}(\theta,\phi)
	\end{array}\right),\;
	\Phi^{(2)}_{\mu, j, m}(\theta,\phi)=\left(\begin{array}{c}
		-\sqrt{\frac{j-m+1}{2j+2}}Y_{\mu,j+\frac{1}{2},m-\frac{1}{2}}(\theta,\phi)\\
		\sqrt{\frac{j+m+1}{2j+2}}Y_{\mu,j+\frac{1}{2},m+\frac{1}{2}}(\theta,\phi)
	\end{array}\right) 
	\ee
	which are eigenvectors of $J^2$ and $J_z$, 
	\be
	J^2 \Phi^{(i)}_{\mu, j, m}=j(j+1)\Phi^{(i)}_{\mu, j, m}\;, \quad J^3 \Phi^{(i)}_{\mu, j, m}=m\Phi^{(i)}_{\mu, j, m}\;, \quad i=1, 2
	\ee	
    Notice that $\Phi^{(2)}$ can be defined for any $j=\mu-\frac{1}{2}, \mu+\frac{1}{2},...$, while $\Phi^{(1)}$ is only defined for $j=\mu+\frac{1}{2}, \mu+\frac{3}{2},...$.
	
	Then one defines $\Omega^{(1)}_{\mu,j,m}$ and $\Omega^{(2)}_{\mu,j,m}$, with $j=\mu+\frac{1}{2}, \mu+3/2,...$, as the following combinations
\bea\nonumber
	& \Omega^{(1)}_{\mu,j,m} = \frac{1}{2}\left(\sqrt{1+\frac{\mu}{j+\frac{1}{2}}}+\sqrt{1-\frac{\mu}{j+\frac{1}{2}}}\right)\Phi^{(1)}_{\mu,j,m}
	-\frac{1}{2}\left(\sqrt{1+\frac{\mu}{j+\frac{1}{2}}}-\sqrt{1-\frac{\mu}{j+\frac{1}{2}}}\right)\Phi^{(2)}_{\mu,j,m}&\\
	\nonumber
	&  \Omega^{(2)}_{\mu,j,m} = \frac{1}{2}\left(\sqrt{1+\frac{\mu}{j+\frac{1}{2}}}-\sqrt{1-\frac{\mu}{j+\frac{1}{2}}}\right)\Phi^{(1)}_{\mu,j,m}
	+\frac{1}{2}\left(\sqrt{1+\frac{\mu}{j+\frac{1}{2}}}+\sqrt{1-\frac{\mu}{j+\frac{1}{2}}}\right)\Phi^{(2)}_{\mu,j,m}		& \\
	\eea
	and 	
	\be
	\Omega^{(3)}_{\mu,\mu-\frac{1}{2},m}= \Phi^{(2)}_{\mu, \mu-\frac{1}{2}, m} \ .
	\ee
	The usefulness of these $\Omega$'s is that they satisfy eq.~(\ref{eq:importanteq})

    \section{Self-adjoint extensions of the Hamiltonian}
    \label{appendix:self-adjoint_extension}

    In this appendix we will solve the time independent Dirac equation for the 2D fermions defined in eq.~(\ref{eq:farmodes}) close to the origin $r=0$, determining the behaviour of the two solutions. We will impose square integrability at the origin, and then impose the condition eq.~(\ref{eq_questioning:self_adjoint_condition}).
    
    One can write down second order differential equations for the two components of $\psi_{2{\rm D}} = (\psi_l,\psi_r)$. For $\psi_{l,r}$ one has
\be
\frac{1}{r}\deriv{(r \psi'_{l,r})}{r} - \left(\frac{\l^2}{r^2}\mp\im \frac{\omega}{r}-\omega^2\right)\psi_{l,r} = 0
\ee

An analytic solution is available in terms of generalized Laguerre polynomials
\be
\label{self-adjoin_app:psiL}
\psi_{l,r} = e^{\mp i\omega r}r^\l\left(c_1 r^{-2\l}L_\l^{-2\l}(\pm 2ir\omega)+c_2L_{-\l}^{2\l}(\pm 2ir\omega)\right)
\ee
These functions are not linearly independent when $\l$ is an integer or half-integer, but all the arguments that will follow are still true by continuity in $\l$.

We now impose square integrability at the origin. We assume $\l>0$ and discuss $\l\le0$ at the end. The crucial observation is that all generalized Laguerre polynomials go to a constant for small $r$, so schematically
\be
\psi_{l,r}\sim c_1 r^{-\l} + c_2 r^\l
\ee
If $c_1\neq0$ we get
\be
\psi_l\sim r^{-\l},\qquad \int_0 |\psi_l|^2 \dd r <\infty\rightarrow \l<\dfrac{1}{2}
\ee
If instead $c_1=0$, $\psi_l$ goes to zero at the origin and integrability is obvious.

Having determined the normalizable states, we now impose their dynamics to be unitary. To achieve this, we impose~(\ref{eq_questioning:self_adjoint_condition}), starting from the special case $\phi=\psi$, from which we get 
\be
\label{eq_self_adjoint:self_adjoint_auto_condition}
(|\psi_l(r)|^2 - |\psi_r(r)|^2)_{r\to 0} = 0
\ee
First, if $\l\ge\frac{1}{2}$ then $c_1=0$ and~(\ref{eq_self_adjoint:self_adjoint_auto_condition}) is trivially satisfied. If $\l<\frac{1}{2}$ $c_1,c_2$ are free and we must more carefully consider $|\psi_l|^2,|\psi_r|^2$. Relying on $0<\l<\frac{1}{2}$ we get
\be
|\psi_l|^2 = |c_1|^2 |L_\l^{-2\l}|^2r^{-2\l} + (\bar{c_1}c_2 \bar{L}_\l^{-2\l}L_{-\l}^{2\l}+\bar{c_2}c_1 L_\l^{-2\l}\bar{L}_{-\l}^{2\l}) + o(1)
\ee
\be
|\psi_r|^2 = |c_1|^2 |L_\l^{-2\l}|^2r^{-2\l} - (\bar{c_1}c_2 \bar{L}_L^{-2\l}L_{-\l}^{2\l}+\bar{c_2}c_1 L_\l^{-2\l}\bar{L}_{-\l}^{2\l}) + o(1)
\ee
so $|\psi_l|^2-|\psi_r|^2\rightarrow0$ if and only if $\bar{c_1}c_2=0$. This means that we actually have two possible phases where the dynamics is unitary: either $c_1=0$ or $c_2=0$. Which one is actually realized depends on the original 4D problem, and might depend on the UV details. Actually, $\l<1/2$ will never arise in our model, but we include it for completeness.
\smallskip

This completes the discussion of the $\l>0$ case. If $\l=0$, the two chiralities decouple and solutions look like
\be
\psi_l = c_l e^{-i\omega r},\qquad \psi_r = c_r e^{i\omega r}
\ee
Substituting $\psi_{l,r}$ into eq.~(\ref{eq_self_adjoint:self_adjoint_auto_condition}) we get
\be
|\psi_l| = |\psi_r|\ \implies \  c_l = e^{i\theta} c_r
\ee
which is the well known supplementary condition that we need to impose for free particles on a half-line. $\theta$ comes from the 4D problem and is UV sensitive.
Finally, $\l<0$ swaps $c_1$ and $c_2$ compared to $\l>0$, but the conclusions are the same.

Focusing now on the relevant cases $\l\ge1/2$ or $\l=0$, we may ask if the full eq.~(\ref{eq_questioning:self_adjoint_condition}) gives any new constraint, and it's easy to see that there are none for $\l\ge1/2$ because there is no free parameter to control. On the other hand, for $\l=0$ we obtain that $\theta$ must be the same for all frequencies $\omega$. Indeed we have
\begin{equation}
    \psi_{l,r} = c_{l,r} e^{\mp i\omega_1 r},\quad c_l = e^{i\theta_1}c_r,\qquad \phi_{l,r} = d_{l,r} e^{\mp i\omega_2 r},\quad d_l = e^{i\theta_2}d_r    
\end{equation}
and condition (\ref{eq_questioning:self_adjoint_condition}) becomes
\begin{equation}
    c_l^*d_l-c_r^*d_r=0 \ \implies \ e^{i\theta_1}=e^{i\theta_2}
\end{equation}
Since the choice of phase is independent of $\omega$, scale symmetry is always preserved. 
		
		\section*{Acknowledgments}
		
		We thank Kenichi Konishi for many discussions and suggestions during the course of this project. The work of S.~B. and  A.~L. is supported by the INFN special research project grant
		``GAST''  (Gauge and String Theories). The work of B.~B.   is supported by the INFN special research project grant
		``TPPC''  (Theoretical Particle Physics and Cosmology).

\appendix

\printbibliography[heading=bibintoc]

@article{Kazama:1976fm,
    author = "Kazama, Yoichi and Yang, Chen Ning and Goldhaber, Alfred S.",
    title = "{Scattering of a Dirac Particle with Charge Ze by a Fixed Magnetic Monopole}",
    reportNumber = "ITP-SB-76/63",
    doi = "10.1103/PhysRevD.15.2287",
    journal = "Phys. Rev. D",
    volume = "15",
    pages = "2287--2299",
    year = "1977"
}

@article{Callan:1982ah,
    author = "Callan, Jr., Curtis G.",
    title = "{Disappearing Dyons}",
    reportNumber = "Print-82-0006 (PRINCETON)",
    doi = "10.1103/PhysRevD.25.2141",
    journal = "Phys. Rev. D",
    volume = "25",
    pages = "2141",
    year = "1982"
}

@article{Cordova:2018cvg,
    author = "C\'ordova, Clay and Dumitrescu, Thomas T. and Intriligator, Kenneth",
    title = "{Exploring 2-Group Global Symmetries}",
    eprint = "1802.04790",
    archivePrefix = "arXiv",
    primaryClass = "hep-th",
    doi = "10.1007/JHEP02(2019)184",
    journal = "JHEP",
    volume = "02",
    pages = "184",
    year = "2019"
}

@article{Callan:1982au,
    author = "Callan, Jr., Curtis G.",
    title = "{Dyon-Fermion Dynamics}",
    reportNumber = "Print-82-0476 (PRINCETON)",
    doi = "10.1103/PhysRevD.26.2058",
    journal = "Phys. Rev. D",
    volume = "26",
    pages = "2058--2068",
    year = "1982"
}

@article{Callan:1982ac,
    author = "Callan, Jr., Curtis G.",
    title = "{Monopole Catalysis of Baryon Decay}",
    reportNumber = "LPTENS-82-20",
    doi = "10.1016/0550-3213(83)90677-6",
    journal = "Nucl. Phys. B",
    volume = "212",
    pages = "391--400",
    year = "1983"
}

@article{Thorngren:2020yht,
    author = "Thorngren, Ryan and Wang, Yifan",
    title = "{Anomalous symmetries end at the boundary}",
    eprint = "2012.15861",
    archivePrefix = "arXiv",
    primaryClass = "hep-th",
    doi = "10.1007/JHEP09(2021)017",
    journal = "JHEP",
    volume = "09",
    pages = "017",
    year = "2021"
}

@article{Rubakov:1982fp,
    author = "Rubakov, V. A.",
    title = "{Adler-Bell-Jackiw Anomaly and Fermion Number Breaking in the Presence of a Magnetic Monopole}",
    doi = "10.1016/0550-3213(82)90034-7",
    journal = "Nucl. Phys. B",
    volume = "203",
    pages = "311--348",
    year = "1982"
}

@online{Csaki:2024ajo,
    author = "Cs\'aki, Csaba and Ovadia, Rotem and Telem, Ofri and Terning, John and Yankielowicz, Shimon",
    title = "{Abelian Instantons and Monopole Scattering}",
    eprint = "2406.13738",
    archivePrefix = "arXiv",
    primaryClass = "hep-th",
    month = "6",
    year = "2024"
}

@article{Jensen:2017eof,
    author = "Jensen, Kristan and Shaverin, Evgeny and Yarom, Amos",
    title = "{\textquoteright{}t Hooft anomalies and boundaries}",
    eprint = "1710.07299",
    archivePrefix = "arXiv",
    primaryClass = "hep-th",
    doi = "10.1007/JHEP01(2018)085",
    journal = "JHEP",
    volume = "01",
    pages = "085",
    year = "2018"
}

@article{Polchinski:1984uw,
    author = "Polchinski, Joseph",
    title = "{Monopole Catalysis: The Fermion Rotor System}",
    reportNumber = "HUTP-84-A005",
    doi = "10.1016/0550-3213(84)90398-5",
    journal = "Nucl. Phys. B",
    volume = "242",
    pages = "345--363",
    year = "1984"
}

@inproceedings{Callan:1983tm,
    author = "Callan, Jr., Curtis g.",
    title = "{THE MONOPOLE CATALYSIS S MATRIX}",
    booktitle = "{Workshop on Problems in Unification and Supergravity}",
    reportNumber = "PRINT-83-0306 (PRINCETON)",
    doi = "10.1063/1.34591",
    pages = "45--53",
    year = "1983"
}

@article{Affleck:1993np,
    author = "Affleck, Ian and Sagi, Jacob",
    title = "{Monopole catalyzed baryon decay: A Boundary conformal field theory approach}",
    eprint = "hep-th/9311056",
    archivePrefix = "arXiv",
    reportNumber = "UBCTP-93-18",
    doi = "10.1016/0550-3213(94)90478-2",
    journal = "Nucl. Phys. B",
    volume = "417",
    pages = "374--402",
    year = "1994"
}

@article{Cordova:2022ieu,
    author = "Cordova, Clay and Ohmori, Kantaro",
    title = "{Noninvertible Chiral Symmetry and Exponential Hierarchies}",
    eprint = "2205.06243",
    archivePrefix = "arXiv",
    primaryClass = "hep-th",
    doi = "10.1103/PhysRevX.13.011034",
    journal = "Phys. Rev. X",
    volume = "13",
    number = "1",
    pages = "011034",
    year = "2023"
}

@article{Sen:1984qe,
    author = "Sen, Ashoke",
    title = "{Role of Conservation Laws in the Callan-Rubakov Process with Arbitrary Number of Generation of Fermions}",
    reportNumber = "FERMILAB-PUB-84-028-T",
    doi = "10.1103/PhysRevLett.52.1755",
    journal = "Phys. Rev. Lett.",
    volume = "52",
    pages = "1755",
    year = "1984"
}

@book{Shnir:2005vvi,
    author = "Shnir, Yakov M.",
    title = "{Magnetic Monopoles}",
    doi = "10.1007/3-540-29082-6",
    isbn = "978-3-540-25277-1, 978-3-540-29082-7",
    publisher = "Springer",
    address = "Berlin/Heidelberg",
    series = "Text and Monographs in Physics",
    year = "2005"
}

@article{Csaki:2022qtz,
    author = "Cs\'aki, Csaba and Shirman, Yuri and Telem, Ofri and Terning, John",
    title = "{Pairwise Multiparticle States and the Monopole Unitarity Puzzle}",
    doi = "10.1103/PhysRevLett.129.181601",
    journal = "Phys. Rev. Lett.",
    volume = "129",
    number = "18",
    pages = "181601",
    year = "2022"
}

@article{Sheu:2022odl,
    author = "Sheu, Chao-Hsiang and Shifman, Mikhail",
    title = "{Consistency of chiral symmetry breaking in chiral Yang-Mills theory with adiabatic continuity}",
    eprint = "2212.14794",
    archivePrefix = "arXiv",
    primaryClass = "hep-th",
    reportNumber = "FTPI-MINN-22-37, UMN-TH-4140/22",
    doi = "10.1103/PhysRevD.107.054030",
    journal = "Phys. Rev. D",
    volume = "107",
    number = "5",
    pages = "054030",
    year = "2023"
}

@article{Csaki:2020yei,
    author = "Cs\'aki, Csaba and Hong, Sungwoo and Shirman, Yuri and Telem, Ofri and Terning, John",
    title = "{Completing Multiparticle Representations of the Poincar\'e Group}",
    eprint = "2010.13794",
    archivePrefix = "arXiv",
    primaryClass = "hep-th",
    doi = "10.1103/PhysRevLett.127.041601",
    journal = "Phys. Rev. Lett.",
    volume = "127",
    number = "4",
    pages = "041601",
    year = "2021"
}

@article{Bolognesi:2022beq,
    author = "Bolognesi, Stefano and Konishi, Kenichi and Luzio, Andrea",
    title = "{Dynamical Abelianization and anomalies in chiral gauge theories}",
    eprint = "2206.00538",
    archivePrefix = "arXiv",
    primaryClass = "hep-th",
    doi = "10.1007/JHEP12(2022)110",
    journal = "JHEP",
    volume = "12",
    pages = "110",
    year = "2022"
}

@article{Csaki:2020inw,
    author = "Csaki, Csaba and Hong, Sungwoo and Shirman, Yuri and Telem, Ofri and Terning, John and Waterbury, Michael",
    title = "{Scattering amplitudes for monopoles: pairwise little group and pairwise helicity}",
    eprint = "2009.14213",
    archivePrefix = "arXiv",
    primaryClass = "hep-th",
    doi = "10.1007/JHEP08(2021)029",
    journal = "JHEP",
    volume = "08",
    pages = "029",
    year = "2021"
}

@article{Bolognesi:2017pek,
    author = "Bolognesi, Stefano and Konishi, Kenichi and Shifman, Mikhail",
    title = "{Patterns of symmetry breaking in chiral QCD}",
    eprint = "1712.04814",
    archivePrefix = "arXiv",
    primaryClass = "hep-th",
    reportNumber = "FTPI-MINN-17-23, UMN-TH-3707-17",
    doi = "10.1103/PhysRevD.97.094007",
    journal = "Phys. Rev. D",
    volume = "97",
    number = "9",
    pages = "094007",
    year = "2018"
}

@article{Csaki:2021ozp,
    author = "Cs\'aki, Csaba and Shirman, Yuri and Telem, Ofri and Terning, John",
    title = "{Monopoles Entangle Fermions}",
    eprint = "2109.01145",
    archivePrefix = "arXiv",
    primaryClass = "hep-th",
    month = "9",
    year = "2021"
}

@article{Hamada:2022eiv,
    author = "Hamada, Yuta and Kitahara, Teppei and Sato, Yoshiki",
    title = "{Monopole-fermion scattering and varying Fock space}",
    eprint = "2208.01052",
    archivePrefix = "arXiv",
    primaryClass = "hep-th",
    reportNumber = "KEK-TH-2423, YITP-22-44, TU-1156",
    doi = "10.1007/JHEP11(2022)116",
    journal = "JHEP",
    volume = "11",
    pages = "116",
    year = "2022"
}

@article{Kitano:2021pwt,
    author = "Kitano, Ryuichiro and Matsudo, Ryutaro",
    title = "{Missing final state puzzle in the monopole-fermion scattering}",
    eprint = "2103.13639",
    archivePrefix = "arXiv",
    primaryClass = "hep-th",
    reportNumber = "KEK-TH-2313, KEK-TH-2321",
    doi = "10.1016/j.physletb.2022.137271",
    journal = "Phys. Lett. B",
    volume = "832",
    pages = "137271",
    year = "2022"
}

@article{vanBeest:2023dbu,
    author = "van Beest, Marieke and Boyle Smith, Philip and Delmastro, Diego and Komargodski, Zohar and Tong, David",
    title = "{Monopoles, Scattering, and Generalized Symmetries}",
    eprint = "2306.07318",
    archivePrefix = "arXiv",
    primaryClass = "hep-th",
    month = "6",
    year = "2023"
}

@article{Armoni:2012xa,
    author = "Armoni, A. and Shifman, M.",
    title = "{A Chiral SU(N) Gauge Theory Planar Equivalent to Super-Yang-Mills}",
    eprint = "1202.1657",
    archivePrefix = "arXiv",
    primaryClass = "hep-th",
    doi = "10.1103/PhysRevD.85.105003",
    journal = "Phys. Rev. D",
    volume = "85",
    pages = "105003",
    year = "2012"
}

@online{vanBeest:2023mbs,
    author = "van Beest, Marieke and Boyle Smith, Philip and Delmastro, Diego and Mouland, Rishi and Tong, David",
    title = "{Fermion-Monopole Scattering in the Standard Model}",
    eprint = "2312.17746",
    archivePrefix = "arXiv",
    primaryClass = "hep-th",
    month = "12",
    year = "2023"
}

@article{Eichten:1985fs,
    author = "Eichten, Estia and Peccei, Roberto D. and Preskill, John and Zeppenfeld, Dieter",
    title = "{Chiral Gauge Theories in the 1/n Expansion}",
    reportNumber = "FERMILAB-PUB-85-147-T, CALT-68-1303",
    doi = "10.1016/0550-3213(86)90206-3",
    journal = "Nucl. Phys. B",
    volume = "268",
    pages = "161--178",
    year = "1986"
}

@article{Khoze:2023kiu,
    author = "Khoze, Valentin V.",
    title = "{Scattering amplitudes of fermions on monopoles}",
    eprint = "2308.09401",
    archivePrefix = "arXiv",
    primaryClass = "hep-th",
    reportNumber = "IPPP/23/XX",
    doi = "10.1007/JHEP11(2023)214",
    journal = "JHEP",
    volume = "11",
    pages = "214",
    year = "2023"
}

@article{Goity:1985tf,
    author = "Goity, J. and Peccei, R. D. and Zeppenfeld, D.",
    title = "{Tumbling and Complementarity in a Chiral Gauge Theory}",
    reportNumber = "DESY-85-051",
    doi = "10.1016/0550-3213(85)90065-3",
    journal = "Nucl. Phys. B",
    volume = "262",
    pages = "95--106",
    year = "1985"
}

@article{Brennan:2021ewu,
    author = "Brennan, T. Daniel",
    title = "{Callan-Rubakov effect and higher charge monopoles}",
    eprint = "2109.11207",
    archivePrefix = "arXiv",
    primaryClass = "hep-th",
    doi = "10.1007/JHEP02(2023)159",
    journal = "JHEP",
    volume = "02",
    pages = "159",
    year = "2023"
}

@online{Brennan:2023tae,
    author = "Brennan, T. Daniel",
    title = "{A New Solution to the Callan Rubakov Effect}",
    eprint = "2309.00680",
    archivePrefix = "arXiv",
    primaryClass = "hep-th",
    month = "9",
    year = "2023"
}

@article{Smith:2019jnh,
    author = "Smith, Philip Boyle and Tong, David",
    title = "{Boundary States for Chiral Symmetries in Two Dimensions}",
    eprint = "1912.01602",
    archivePrefix = "arXiv",
    primaryClass = "hep-th",
    doi = "10.1007/JHEP09(2020)018",
    journal = "JHEP",
    volume = "09",
    pages = "018",
    year = "2020"
}

@article{Harvey:1983tp,
    author = "Harvey, Jeffrey A.",
    title = "{Magnetic Monopoles With Fractional Charges}",
    reportNumber = "Print-83-0570 (PRINCETON)",
    doi = "10.1016/0370-2693(83)91101-2",
    journal = "Phys. Lett. B",
    volume = "131",
    pages = "104--110",
    year = "1983"
}

@online{Smith:2020nuf,
    author = "Smith, Philip Boyle and Tong, David",
    title = "{What Symmetries are Preserved by a Fermion Boundary State?}",
    eprint = "2006.07369",
    archivePrefix = "arXiv",
    primaryClass = "hep-th",
    month = "6",
    year = "2020"
}

@article{Rubakov:1988aq,
    author = "Rubakov, V. A.",
    title = "{Monopole Catalysis of Proton Decay}",
    doi = "10.1088/0034-4885/51/2/002",
    journal = "Rept. Prog. Phys.",
    volume = "51",
    pages = "189--241",
    year = "1988"
}

\end{document}